\documentclass[12pt,preprint]{aastex}
\shorttitle{ULYSSES observations of the polar solar wind}
\shortauthors{R. M. Nicol, S. C. Chapman and R. O. Dendy}
\begin{document}
\bibliographystyle{plainnat}
\title{THE SIGNATURE OF EVOLVING TURBULENCE IN QUIET SOLAR WIND AS SEEN BY ULYSSES}
\author{R. M. N\fontshape{sc}\selectfont{icol} \altaffilmark{1}, S. C. C\fontshape{sc}\selectfont{hapman} \altaffilmark{1} and R. O. D\fontshape{sc}\selectfont{endy} \altaffilmark{1} \altaffilmark{2}}
\altaffiltext{1}{Centre for Fusion, Space and Astrophysics, Department of Physics, University of Warwick, Coventry, CV4 7AL, UK}
\email{R.M.Nicol@warwick.ac.uk}
\altaffiltext{2}{UKAEA Culham Division, Culham Science Centre, Abingdon, Oxfordshire, OX14 3DB, UK}
\begin{abstract}
Solar wind fluctuations, such as magnetic field or velocity, show power law power spectra suggestive both of an inertial range of intermittent turbulence (with $\sim -5/3$ exponent) and at lower frequencies, of fluctuations of coronal origin (with $\sim -1$ exponent).
The ULYSSES spacecraft spent many months in the quiet fast solar wind above the Sun's polar coronal holes in a highly ordered magnetic field. We use statistical analysis methods such as the generalized structure function (GSF) and extended self-similarity (ESS) to quantify the scaling of the moments of the probability density function of fluctuations in the magnetic field.
The GSFs give power law scaling in the ``$f^{-1}$'' range of the form $\langle\vert y(t+\tau)-y(t)\vert^{m}\rangle\sim\tau^{\zeta(m)}$, but ESS is required to reveal scaling in the inertial range, which is of the form $\langle\vert y(t+\tau)-y(t)\vert^{m}\rangle\sim [g(\tau)]^{\zeta(m)}$.
We find that $g(\tau)$ is independent of spacecraft position and $g(\tau)\sim\tau^{-log_{10}(\tilde{\lambda}\tau)}$. The ``$f^{-1}$'' scaling fluctuates with radial spacecraft position. This confirms that, whereas the ``$f^{-1}$'' fluctuations are directly influenced by the corona, the inertial range fluctuations are consistent with locally evolving turbulence, but with an ``envelope'' $g(\tau)$, which captures the formation of the quiet fast solar wind.
\end{abstract}
\section{Introduction}
The solar wind provides a unique opportunity to perform in-situ long interval studies of a high magnetic Reynolds number \citep{Matt2point}, magnetohydrodynamic stellar wind. Magnetic field fluctuations in the solar wind typically exhibit an extended power law region in the power spectrum \citep[e.g,][]{kolmogorov68,livingreview}, from timescales of a few hours down to that characteristic of the ion dynamics. Solar wind acceleration and heating to form the fast solar wind is known to occur on open flux lines, i.e. over coronal holes (see \citet{coronalheating} for a recent review). The solar polar coronal holes thus provide an uninterrupted, spatially extended region in which we can study the accelerated fast solar wind.

A key objective of the present work is to relate aspects of the ULYSSES solar wind measurements described in the next section: their spectral power density; their intermittency \citep{radialintermittency}; and their spatial location. Generalizing somewhat, the solar wind spectral power density is observed to scale approximately as ``$f^{-1}$'' \citep{confAIP,reviewMHDinSW} at lower frequencies ($\leq1$ mHz); and as ``$f^{-5/3}$'' \citep{fractalRuz,secminKol}, reminiscent of the inertial range of Kolmogorov \citep{K41}, at higher frequencies ($\sim10$ mHz-$100$ mHz). The frequency at which the transition occurs ($\sim1$ mHz-$10$ mHz) between these two power laws is observed to decline with increasing distance from the sun in the plane of the ecliptic \citep[see][]{breakpoint,turbulentevolution}. This extension of the ``$f^{-5/3}$'' range to lower frequency at greater distances can be interpreted as evidence for an active turbulent cascade \citep{reviewMHDinSW,MHDcascade97} that evolves a growing inertial range as time passes in the outward propagating plasma. The ``$f^{-1}$'' component is taken to reflect embedded coronal turbulence, convected with the solar wind plasma \citep{PRL1/f86}. It is not yet certain whether, in addition, this coronal turbulence acts as the low frequency large-scale driver of the inertial range turbulence. Both the ``$f^{-1}$'' and ``$f^{-5/3}$'' fluctuations are often predominantly shear Alfv\'enic in character, that is incompressible and displaying correlation or anticorrelation between perturbations of the magnetic field and of fluid velocity \citep{polaralfven}.

The first observations initiated a debate as to whether the fluctuations are of solar coronal origin, simply passively advecting with the flow, or whether they are dominated by locally evolving turbulence \citep{majprobspp}.
The inertial range fluctuations, and the crossover to ``$f^{-1}$'' behavior, show secular variation with heliographic distance \citep{turbulentevolution} consistent with evolving, rather than fully evolved, turbulence. Helios data, in conjunction with ULYSSES, has been used by \citet{mhd_high_lat}, to show that the ``$f^{-1}$'' region contains scales, which are too large to be produced in situ and are progressively eaten away by the smaller scale turbulence, which must therefore be active.
The large scale magnetic structure of the corona also varies with both heliospheric latitude and solar cycle, and this is clearly manifested in the coherent structures and variation of wind speed that are observed \citep{plasmaobservations}. Power spectra are not sufficient to quantify fully the scaling properties of fluctuations \citep[e. g.][]{scaling05}.

Recent structure function analysis using WIND and ACE data, taken in the ecliptic plane at $1$AU, shows evidence of scaling within the inertial range that is solar cycle dependent \citep{selfsimilarwind}. To unravel the interplay between the large scale coronal driver and the evolving inertial range turbulence, we therefore make use of ULYSSES polar passes. ULYSSES' out-of-ecliptic orbit enables the study of possible latitudinal and radial dependences of the solar wind \citep{ulysseslatgradient,turbulentevolution}. In particular, ULYSSES spent many months in the quiet fast solar wind above the polar coronal holes at periods of both minimum and maximum solar activity. The passes during periods of minimum solar activity have been extensively studied because they allow observations of magnetohydrodynamic (MHD) fluctuations free from perturbations from large scale events such as coronal mass ejections (CMEs) or large scale stream structures. The instruments aboard ULYSSES relevant to studies of solar wind fluctuations are the Vector Helium Magnetometer (Jet Propulsion Laboratory) and the Fluxgate Magnetometer (Imperial College) \citep{instrumentULYSSES}. Plasma flow measurements are obtained from the SWOOPS experiment \citep{SWOOPS}, and the time resolutions available range from seconds for the magnetic field measurements to $4$ minute averages for the electron and ion velocities. Thus the magnetometer data offers a sufficiently extensive dynamical range to explore the inertial range. ULYSSES has performed five polar passes and previous studies include \citet{invariantBr,magsignature,Evolutionmag} and references therein.

In this paper, we seek to characterize the intermittency therefore we remove the minimum of outliers consistent with obtaining good representation of the tails of the probability density functions (PDFs) \citep{scalingmethod06}. This is distinct from, but complimentary to, approaches that seek to eliminate the intermittency by removing significant fractions of the tails of the PDFs, in order to probe the remnant scaling (S. Habal, private communication, 2007). We perform generalized structure function analysis (GSF) \citep{Frisch}, on intervals of quiet solar wind as seen by ULYSSES. We show that while GSF is not sufficient to reveal scaling in the inertial range, extended self-similarity (ESS) \citep{ESS93,ESSCarbone}, successfully recovers self-similar behavior. Furthermore we examine the possible latitudinal and radial dependences of both the inertial and ``$f^{-1}$'' ranges, and conclude that the inertial range, unlike the ``$f^{-1}$'' regime, shows very little variation with spacecraft position over the $60$ days considered. We focus here on magnetic field fluctuations for our analysis rather than Els\"{a}sser variables, since these involve velocity measurements at lower cadence; however see \citet{PRLSorriso07}, where local (in time) scaling properties are also considered.

\section{Quantifying Scaling}
We first introduce the analysis tools that enable us to quantify scaling in a timeseries. Generally speaking, a time series $y(t)$ exhibits scaling \citep{Sornette}, if
\begin{equation}
\langle\vert y(t+\tau)-y(t)\vert^{m}\rangle\sim\tau^{\zeta(m)}\label{eqn1}
\end{equation}
Here the angular brackets denote an ensemble average over $t$, implying an assumption of approximate stationarity. In practice, we examine the data over a sufficiently large range of time intervals $\tau$ in order to establish the power law dependence (see equation (\ref{eqn1})); that is, the scaling exponent $\zeta(m)$.

The point of contact between equation (\ref{eqn1}) and fluid turbulence is the Taylor hypothesis \citep{Taylor38}, where in the high speed flow the time interval $\tau$ plays the role of a longitudinal lengthscale. However equation (\ref{eqn1}) expresses a generic scaling property, which is also found, for example, in random fractals such as Brownian walks and L\'{e}vy flights \citep{Sornette}. Since the coronal magnetic carpet is known to be fractal \citep{mag_carpet_2} and the large scale coronal dynamics as seen in solar flare statistics exhibit scaling \citep{solarflare}, one might anticipate that a propagating signature of coronal origin might also show scaling which could be captured by equation (\ref{eqn1}). In practice, we test equation (\ref{eqn1}) by computing the associated generalized structure functions or GSF \citep[][and references therein]{fractalBurlaga,fractalRuz,intermittentturbsouth,spatialspectralexponent,MHDcascade97}:
\begin{equation}
S_m(\tau)=\langle\vert y(t+\tau)-y(t)\vert^{m}\rangle=\langle\vert \delta y\vert^{m}\rangle\label{eqn2}
\end{equation}
For a perfectly self-similar process, such as the Kolmogorov cascade or a random fractal, $\zeta(m)$ depends linearly on $m$. Turbulent flows are however typically intermittent, that is, the dissipation does not occur uniformly in space or time. The corresponding exponents $\zeta(m)$ are quadratic in $m$ \citep{Frisch}. Nevertheless, $S_m$ scaling with $\tau$ would be expected in uniform, fully developed turbulence in an infinite domain. 

In practice, both in numerical simulations \citep{simulmerrifield} and in the laboratory \citep{L_H_mast,characternonlinear} the scaling in equation (\ref{eqn1}) is not always found. This may reflect the finite spatial domain, or that the turbulence is not fully evolved. We will also see that this is the case for the inertial range in quiet intervals of ULYSSES observations. However a weaker form of scaling, known as extended self-similarity (ESS) turns out to be applicable to our datasets in cases where equation (\ref{eqn1}) is not. ESS proceeds by replacing $\tau$ in equation (\ref{eqn1}) by an initially unknown generalized timescale $g(\tau)$, such that formally
\begin{equation}
S_m(\tau)\sim[g(\tau)]^{\zeta(m)}\label{eqn3}
\end{equation}
It follows from equation (\ref{eqn3}) that
\begin{equation}
S_m(\tau)=[S_{m'}(\tau)]^{\zeta(m)/\zeta(m')}\label{eqn4}
\end{equation}
\citep[see for example][]{ESSGross,magstudyUlysses}.

The measured vector magnetic field time series, $\mathbf{B}(t)$, is differenced for time lags $\tau$ in the range $1$ minute to $50$ minutes, yielding a series $\delta y_i(t,\tau)$ for its three components
\begin{equation}
\delta y_i(t,\tau)=B_i(t+\tau)-B_i(t)\label{eqn5}
\end{equation}
where $i$ denotes the component of $\mathbf{B}$ under consideration.
Assuming time-stationarity, the $t$ dependence in $\delta y_i(t,\tau)$ can be dropped:
\begin{equation}
S_{m}(\tau)=\langle\vert \delta y_i\vert^{m}\rangle=\int^{\infty}_{-\infty}\vert\delta y_i\vert^{m}P(\delta y_i,\tau)d(\delta y_i)\label{eqn6}
\end{equation}
where $P(\delta y_i,\tau)$ is the probability density function of $\delta y_i$, $<\cdot>$ again denotes temporal averaging and $m$ is a positive integer. The effect of outliers becomes increasingly apparent in the higher order structure functions, subject to statistical constraints, which typically limits consideration to $m\leq 6$. One of the main problems associated with the use of structure functions is the nature of the limits in equation (\ref{eqn6}) \citep{scaling05}. For this reason a clipping technique developed by \citet{scalingmethod06} is applied, when indicated below, to condition the data. This involves removing an increasing fraction of the maximum and minimum $y_i$ values and observing the effect on the scaling. Both raw datasets and conditioned datasets will be shown.

\section{The datasets}
In 1995, the ULYSSES spacecraft spent three months above the North polar coronal hole, in quiet fast solar wind. As this was close to a period of minimum solar activity, the magnetic topology of the Sun was relatively simple, free from transient events such as solar flares. The Sun's surface magnetic field was actually dipolar during this time, positive or outwards in the Northern hemisphere and negative or inwards in the Southern hemisphere \citep{solarmin}.

Throughout this paper, we present results separately for each $10$ day contiguous interval from day $180$ to $239$ of $1995$, while ULYSSES was above the Northern coronal hole. Each dataset comprises approximately $13,000$ datapoints of one minute averaged measurements. This enables us to explore both the scaling properties of the inertial and ``$f^{-1}$'' ranges and to test for radial and latitudinal dependencies. Over these $60$ days, ULYSSES moved from $1.7926$ AU to $2.2043$ AU heliospheric distance and through $73.76\ensuremath{^\circ}$ to $77.03\ensuremath{^\circ}$ heliographic latitude, with a peak at $80.22\ensuremath{^\circ}$ on days $212$ and $213$. The successive time intervals are compared in order to identify any radial or latitudinal trend.
The study is restricted to the radial ($R$), tangential ($T$) and normal ($N$) magnetic field components and uses one minute averaged measurements in order to remove any possible sub-spacecraft spin artefacts since ULYSSES has a spin period of $12$s. The $RTN$ coordinate system corresponds to solar-ordered coordinates where $R$ is the sun-ULYSSES axis, $T$ is the cross product of $R$ with the solar rotation axis, and $N$ is the cross product of $R$ and $T$, completing the right-handed system.

\section{ULYSSES observations and scaling}
\subsection{Power Spectra}
\begin{figure}
\figurenum{1}
\epsscale{0.4}
\plotone{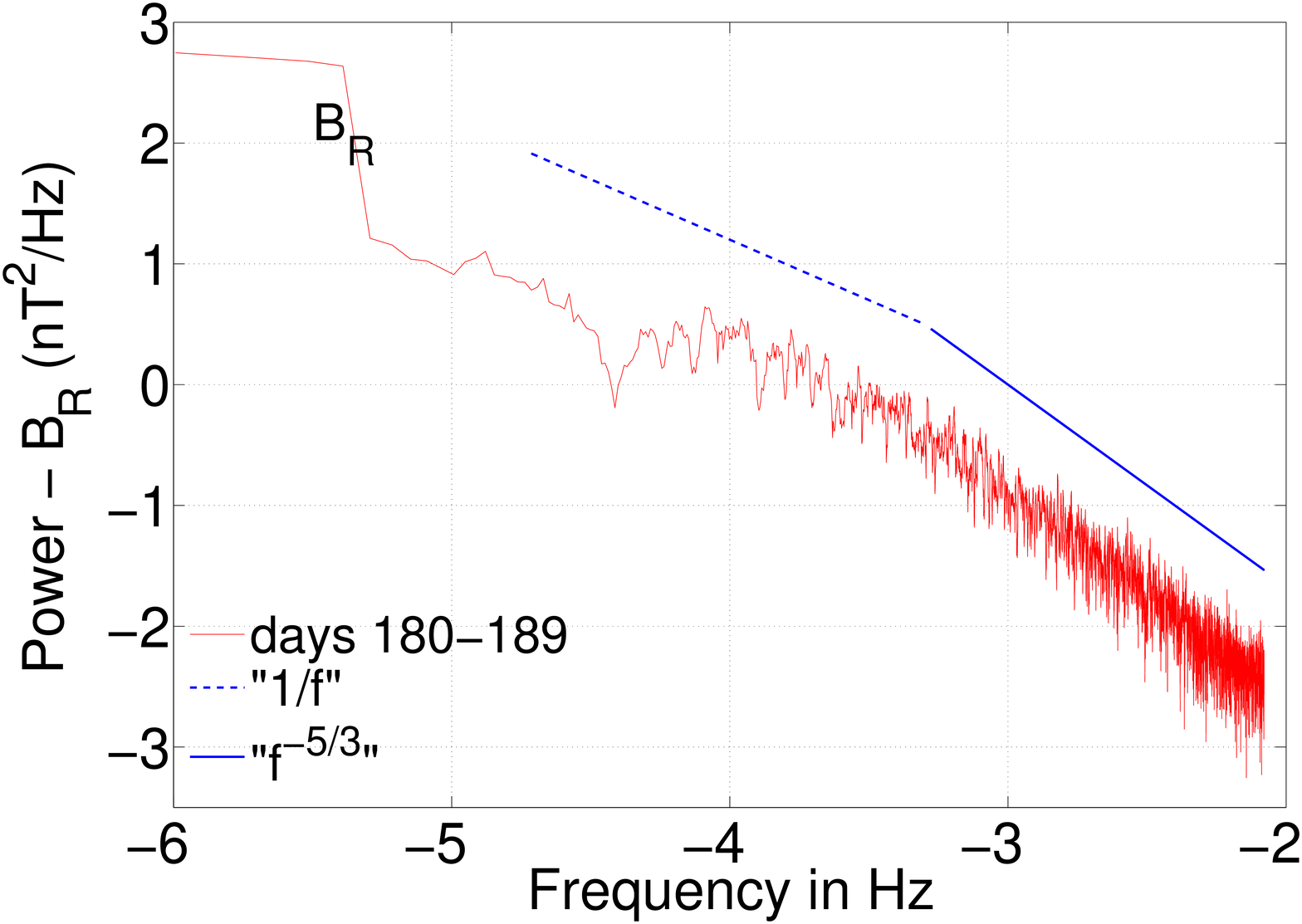}
\plotone{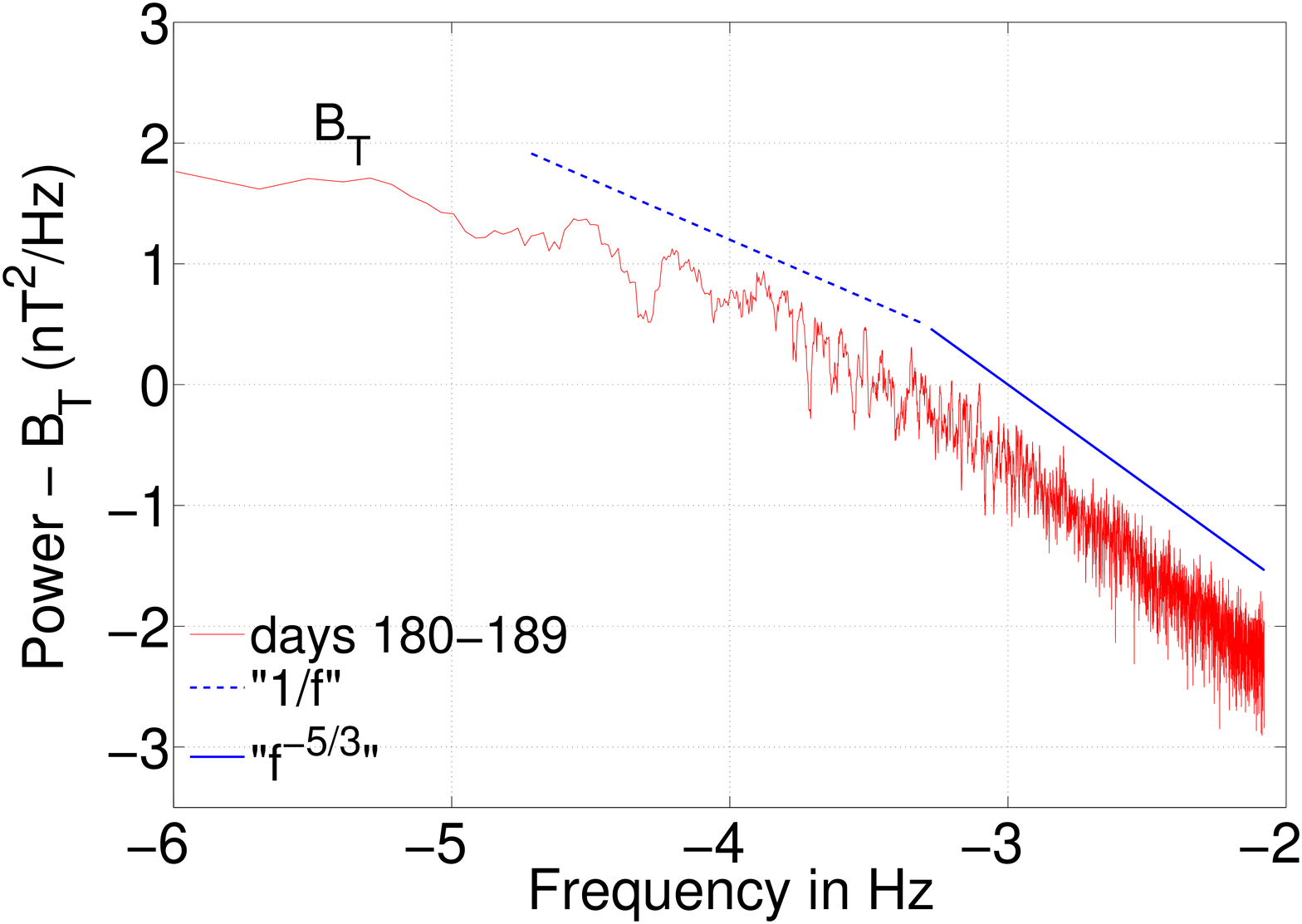}
\plotone{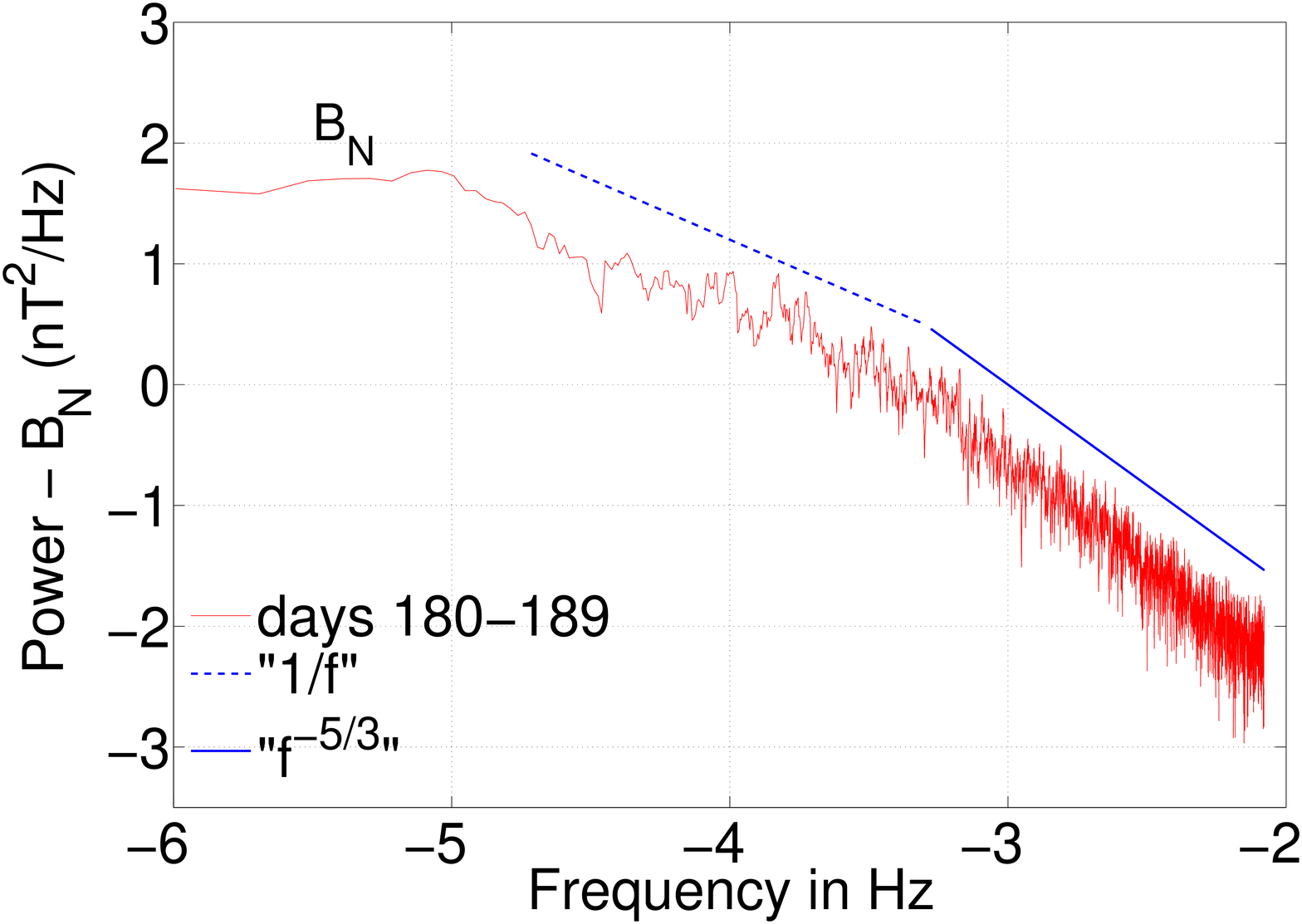}
\caption{Log-log plots of the B-field components' power spectra for days $180-189$. Two regions with different scaling exponents are distinguishable with a break between frequencies at $10^{-3.5}-10^{-3}$Hz, consistent with previous results \citep{turbulentevolution}. For comparison purposes, the $-1$ and $-5/3$ power scaling laws are also shown. The power spectra for the other time intervals examined show similar behavior.}
\label{Fig.1}
\end{figure}

The magnetic field power spectra are computed using the multitapering spectral analysis method \citep{pmtm}. In Figure \ref{Fig.1} we see that the power spectra show an inertial range with a Kolmogorov-like behavior at higher frequencies and a characteristic flattening of the spectra at lower frequencies The existence of this regime is well-known in many physical processes \citep{1overfnoise}, including the interplanetary magnetic field \citep{densityandB1overf}. The power spectra reveal power law scaling, but give no information on intermittency or on whether the turbulent cascade is active; for this we turn to the associated GSFs.
\subsection{Generalized Structure Functions}
\begin{figure}
\figurenum{2}
\epsscale{0.4}
\plotone{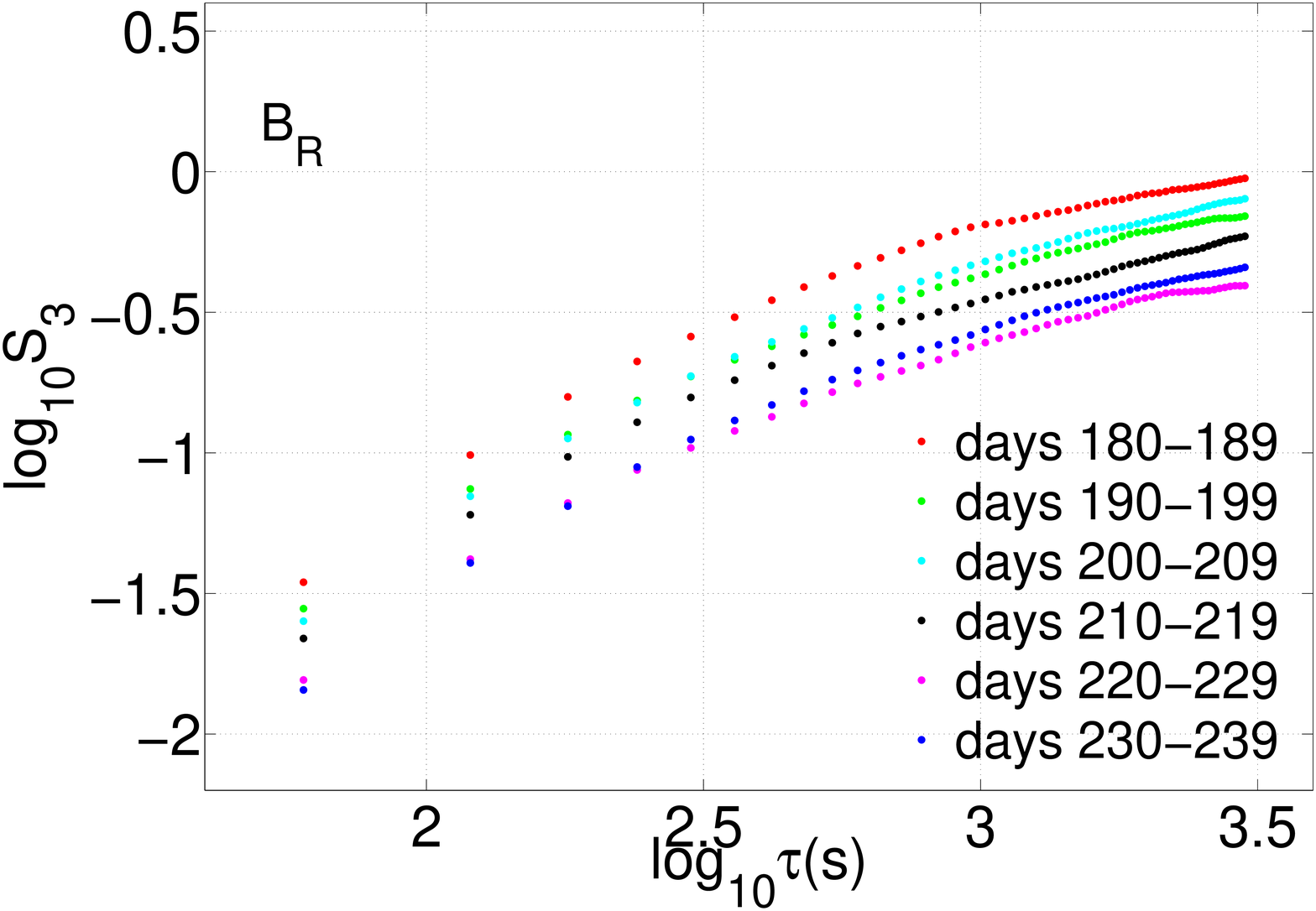}
\plotone{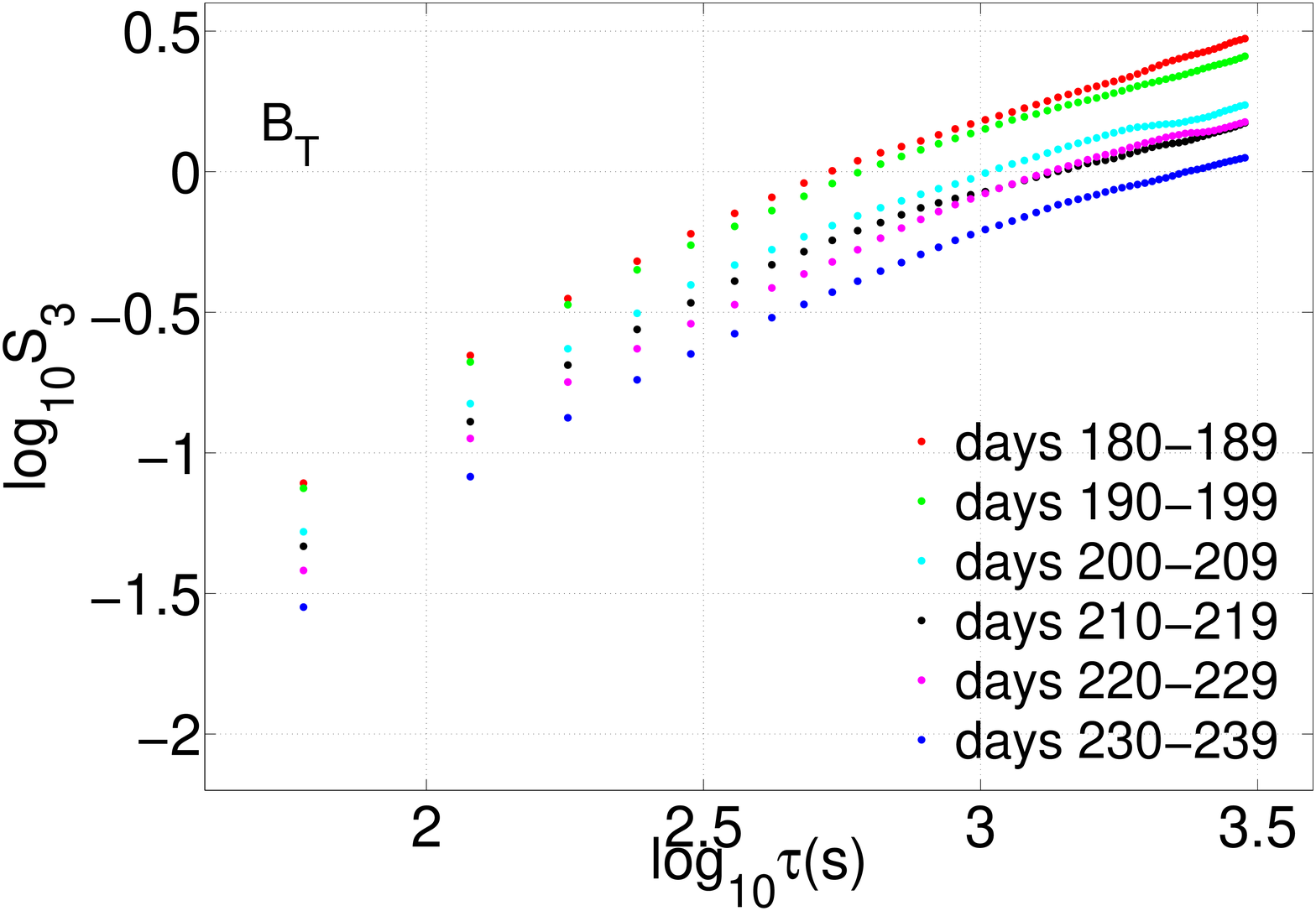}
\plotone{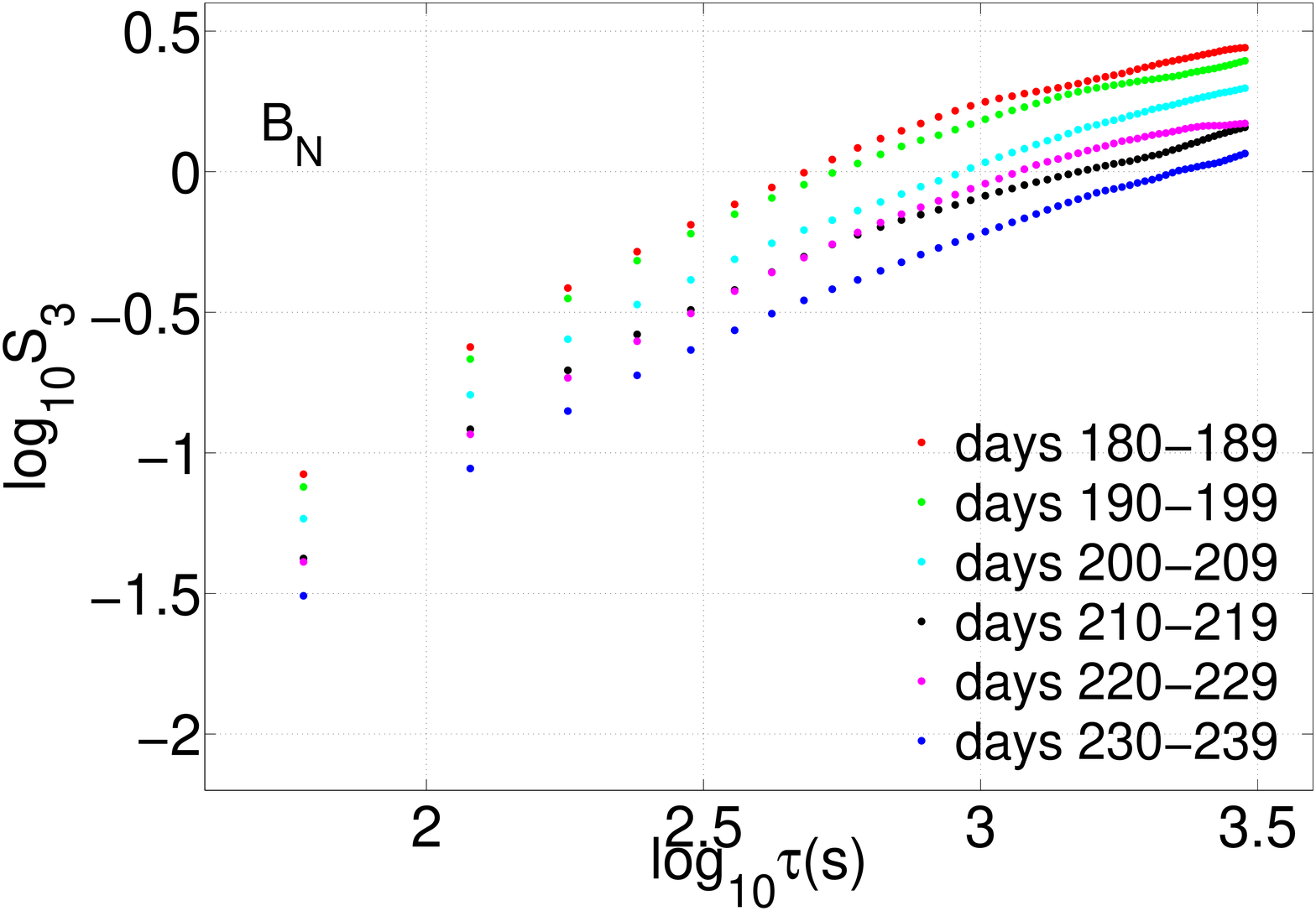}
\caption{Log-log plots of third order structure function $S_3$ versus sampling interval $\tau$ for all three components of magnetic field fluctuations in the solar wind measured by the ULYSSES spacecraft during contiguous intervals of ten days, which are plotted separately on each panel, from day $180$ to day $239$ of $1995$. Only the $0.1$\% conditioned data is shown for clarity. The evolution of the spectral breakpoint to larger $\tau$ with increasing radial distance (or time) can be seen from the change in curvature of $logS_3$ between the different time intervals.}
\label{Fig.2}
\end{figure}

\begin{figure} 
\figurenum{3}
\epsscale{0.4}
\plotone{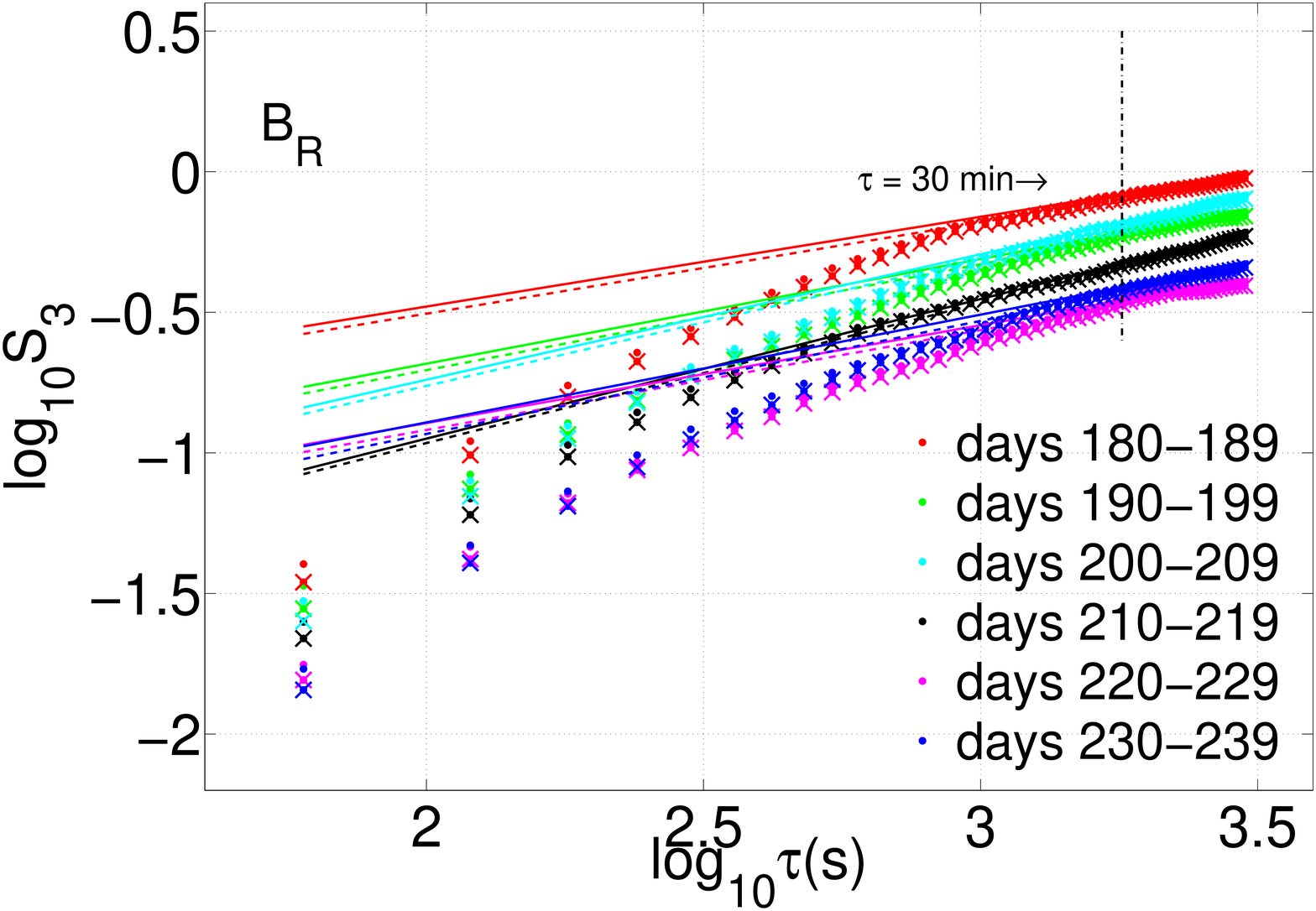}
\plotone{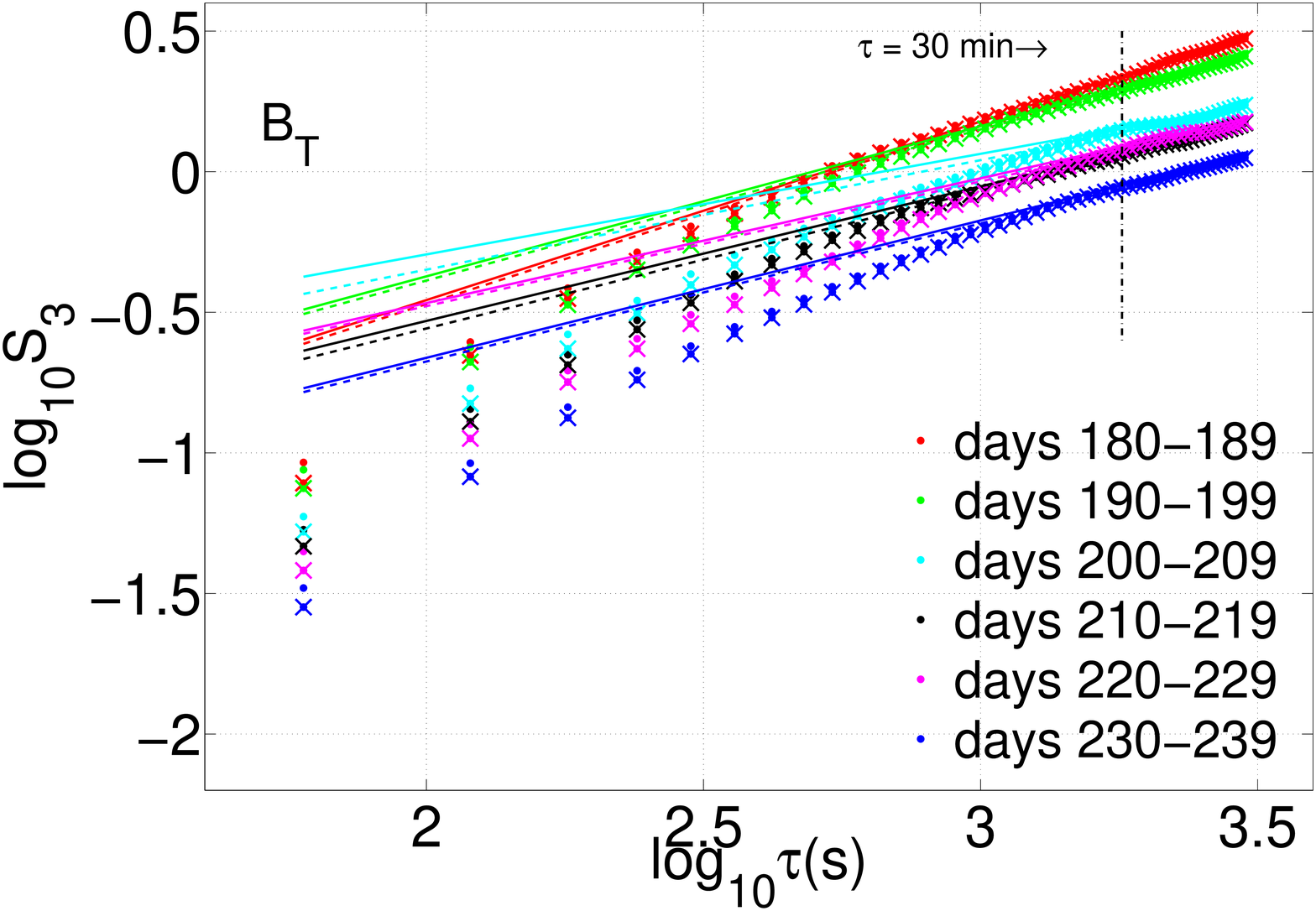}
\plotone{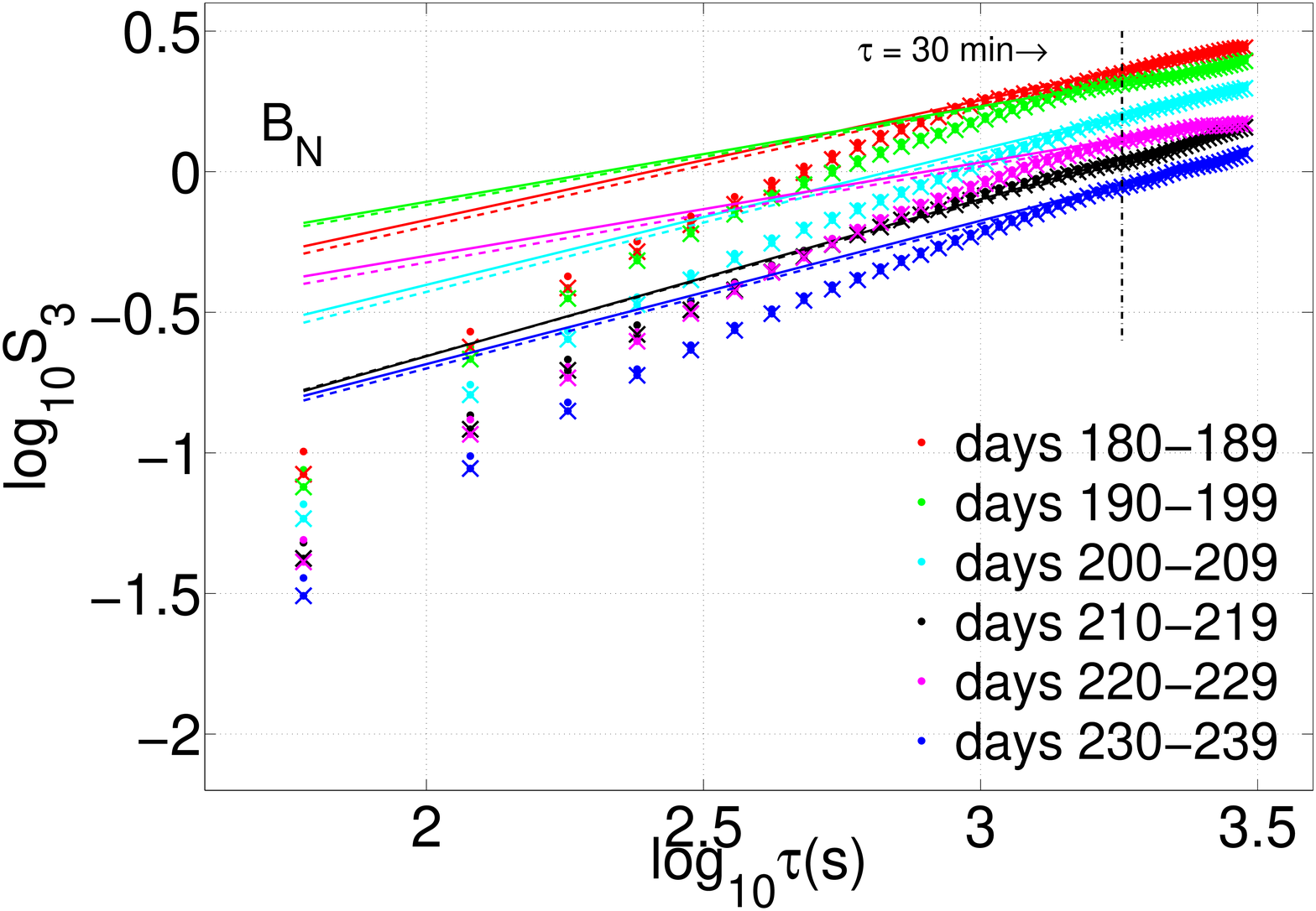}
\caption{Log-log plots of third order structure function $S_3$ versus sampling interval $\tau$ for all three components of magnetic field fluctuations in the solar wind measured by the ULYSSES spacecraft during contiguous intervals of ten days, which are plotted separately on each panel, from day $180$ to day $239$ of $1995$. The raw data is represented by ``\mbox{\large$\cdot$}'' and the $0.1\%$ conditioned data by ``$\times$''. Top left panel: radial field $B_R$. Top right panel: tangential field $B_T$. Bottom panel: normal field $B_N$. For $\tau\geq30$ minutes, corresponding to ``$f^{-1}$''  power spectral density, this gives evidence for scaling of the form equation (\ref{eqn2}) with $\zeta_{L,R}(3) = 0.399\pm0.011$, $\zeta_{L,T}(3) = 0.508\pm0.010$ and $\zeta_{L,N}(3) = 0.445\pm0.008$; solid lines show linear regression fits for $26$ minutes $\geq\tau\geq$ $49$ minutes for $0.1\%$ conditioned data, whereas the dashed lines show the same fits for the raw data. For $\tau\leq30$ minutes, corresponding to the inertial range, there is no scaling but the data suggest a possible common $g(\tau)$ defined by equation (\ref{eqn3}).}
\label{Fig.3}
\end{figure}

We first summarise the scaling behavior seen in the different time intervals in Figure \ref{Fig.2}, where we plot $S_3$ versus $\tau$ for the data conditioned by clipping $0.1\%$ outliers following the technique of \citet{scalingmethod06}. The existence of two distinct scaling regions is clear.
For the small $\tau$ region we see the inertial range with scaling exponents $\zeta_S(m)$, whereas the large $\tau$ region corresponds to the ``$f^{-1}$'' range with scaling exponents $\zeta_L(m)$. Figure \ref{Fig.3} shows that linear regression applied to the third order structure functions for both raw and conditioned data yields power-law scaling in the ``$f^{-1}$'' range. However this does not appear to be the case for the inertial range, where the $S_{m}$ clearly curve for $\tau\leq30$ minutes. The $\zeta$ exponent numerical values for the raw and conditioned GSF data differ by no more than $\sim6\%$ on average, with the higher order moments showing stronger variations, as expected.

\begin{figure} 
\figurenum{4}
\epsscale{0.4}
\plotone{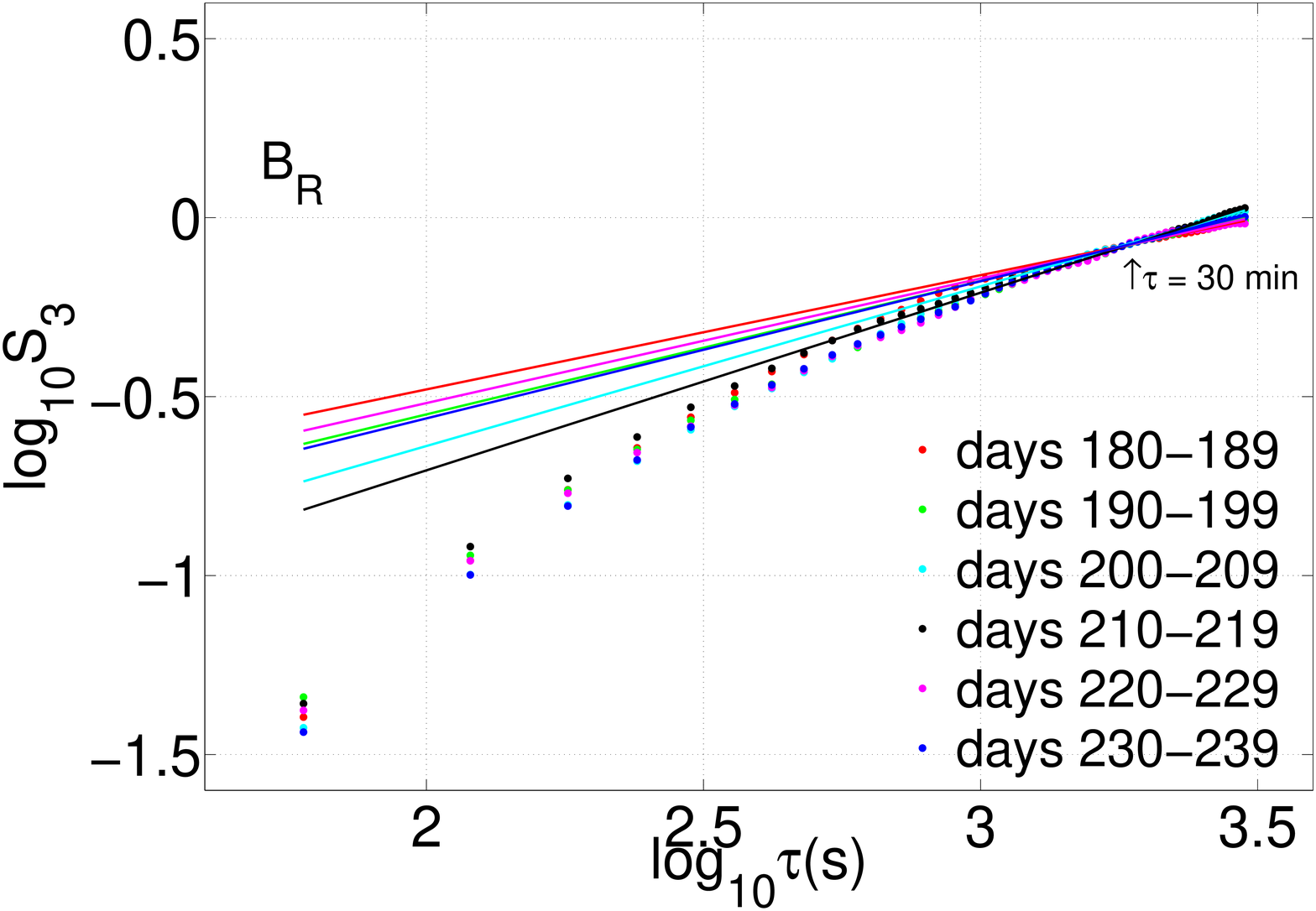}
\plotone{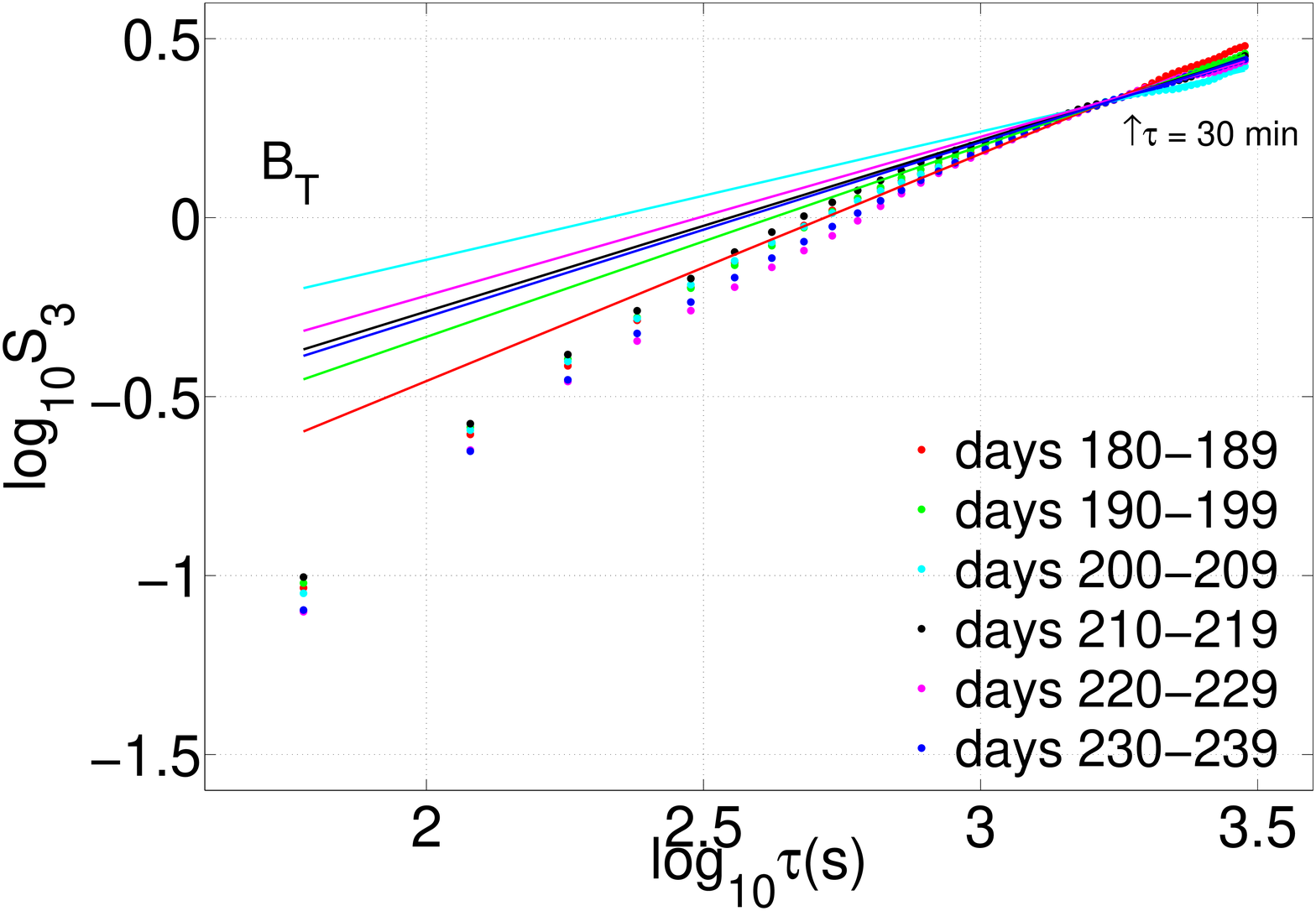}
\plotone{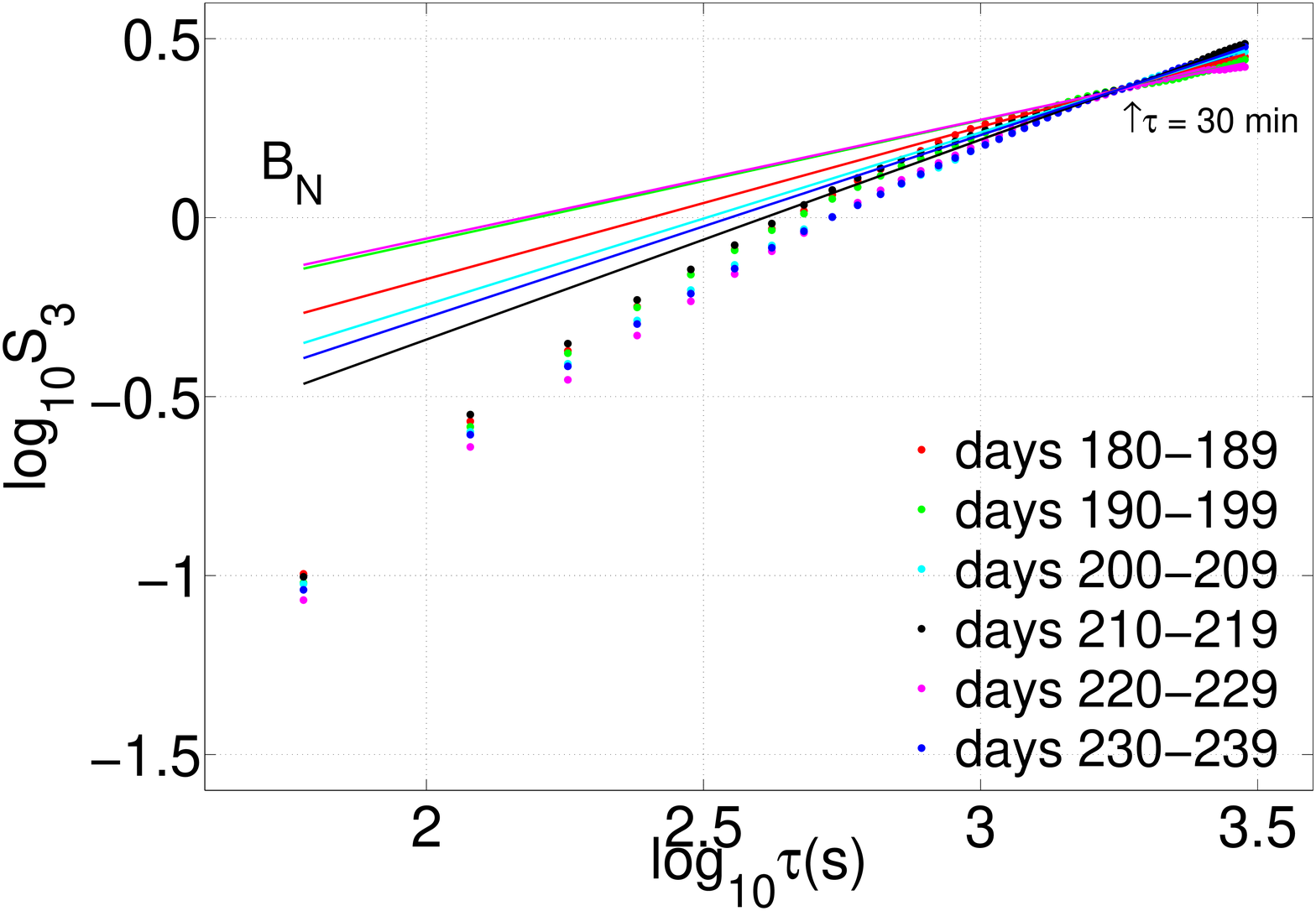}
\caption{Evidence for limited variation in $\zeta(3)$ between ten-day data runs in the scaling region, and for a common $g(\tau)$ between the different time intervals. Data for $B_R$, $B_T$ and $B_N$ from Figure \ref{Fig.2} are replotted after normalization such that the value of $S_3(\tau = 30$ minutes) is the same for each ten-day run within each panel. Gradients in the scaling range (solid lines) do not vary systematically with their timing with respect to the 60-day observation period, nor do they vary in the same way for different magnetic field components. Away from the scaling region, points appear to lie close to a common curve, reflecting $g(\tau)$.}
\label{Fig.4}
\end{figure}

Figures \ref{Fig.2} and \ref{Fig.3} suggest that a single function $g(\tau)$ may be common to all the time intervals considered. We test this conjecture in Figure \ref{Fig.4}, by normalizing all the $S_3$ log plots to their value at $\tau=30$min., close to the centre of the $\tau$ range. The curves overlay quite closely, and there is no secular latitudinal or radial dependence of $g(\tau)$.
The displacements of the curves in Figures \ref{Fig.2} and \ref{Fig.3} arise from the data: as ULYSSES moves, the GSF and ESS plots shift in a relatively ordered way. In Figure \ref{Fig.5} we examine this further by showing the variations of a single point on each line, $S_3(\tau=30)$ as a function of time and therefore of increasing radial distance, for all $B$ field components. $S_3(\tau=30)$ for all three components exhibits a secular trend and decreases with time as heliographic range increases. There is no significant latitudinal variation, although there is a flattening of the slope after $\sim$ day $210$ (or interval $4$), when ULYSSES passes over the solar pole and the heliographic latitude starts to decrease again. This is more apparent in the $N$ and $T$ components, the $R$ component seeming relatively unaffected. This is consistent with the work of \citet{mhd_high_lat}, who also observed a stronger radial, rather than latitudinal, dependence of the turbulent properties.
\subsection{Extended self-similarity}
Let us now test more precisely for $S_{m}\sim [g(\tau)]^{\zeta(m)}$ within the inertial range by applying ESS to the data. We first apply the technique to the entire $\tau$ range of $2-49$ min., with results shown in Figures \ref{Fig.6} and \ref{Fig.7}.

Power law scaling is recovered for the entire $\tau$ range considered.
ESS also seems to extend the inertial range scaling region and, apart from a few exceptional cases, it is generally difficult to distinguish clearly between the inertial range and ``$f^{-1}$'' ranges from these plots. Figure \ref{Fig.6} shows a reasonable linear fit to the whole $\tau$ range but we see a small but systematic displacement of the data from the fitted line at higher frequencies. Figure \ref{Fig.7} shows the ESS exponents $\zeta(2)/\zeta(3)$, obtained from the gradients of the linear fits in Figure \ref{Fig.6}, which exhibit no dependence on the time interval considered. 

However the GSF analysis has given prior indication of the $\tau$ at which the transition from one regime to the other occurs. It is therefore of interest to apply ESS analysis separately to the two regions - inertial range and ``$f^{-1}$'' - identified above. As we discuss below, Figures \ref{Fig.8} to \ref{Fig.10} then demonstrate that the ``$f^{-1}$'' range is the dominant source of variation, in clear contrast to the inertial range. Figure \ref{Fig.8} repeats the ESS analysis for the inertial range with $\tau=2-14$ min., and Figure \ref{Fig.9} for the ``$f^{-1}$'' range with $\tau=26-49$ min.

\begin{figure}
\figurenum{5}
\epsscale{0.8}
\plotone{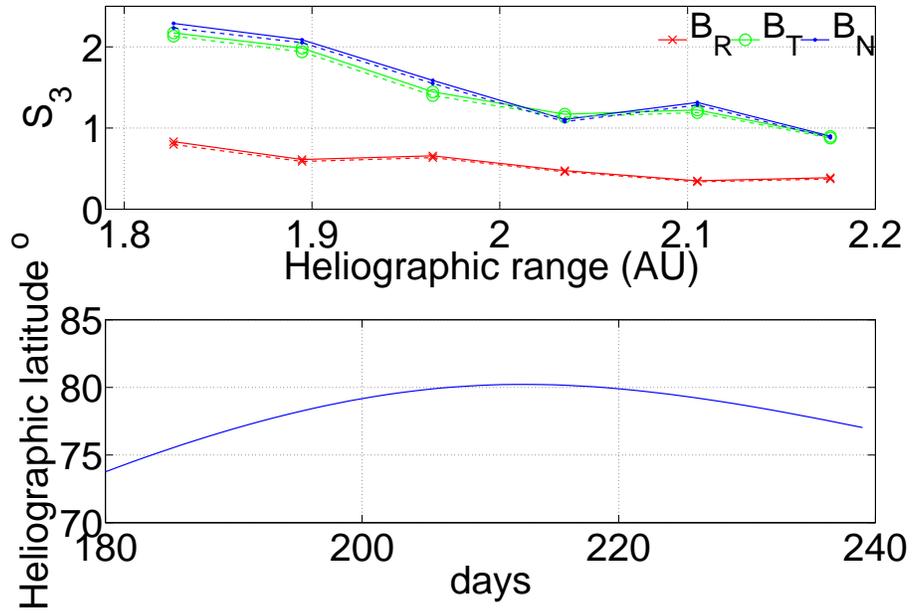}
\caption{Dependence of the value of $S_3(\tau = 30$ minutes) on sampling time - a proxy for mean location of the moving ULYSSES spacecraft - for $B_R$, $B_T$ and $B_N$. Here interval one refers to days $180$ to $189$ of $1995$, and interval six to days $230$ to $239$.}
\label{Fig.5}
\end{figure}

\begin{figure} 
\figurenum{6}
\epsscale{0.4}
\plotone{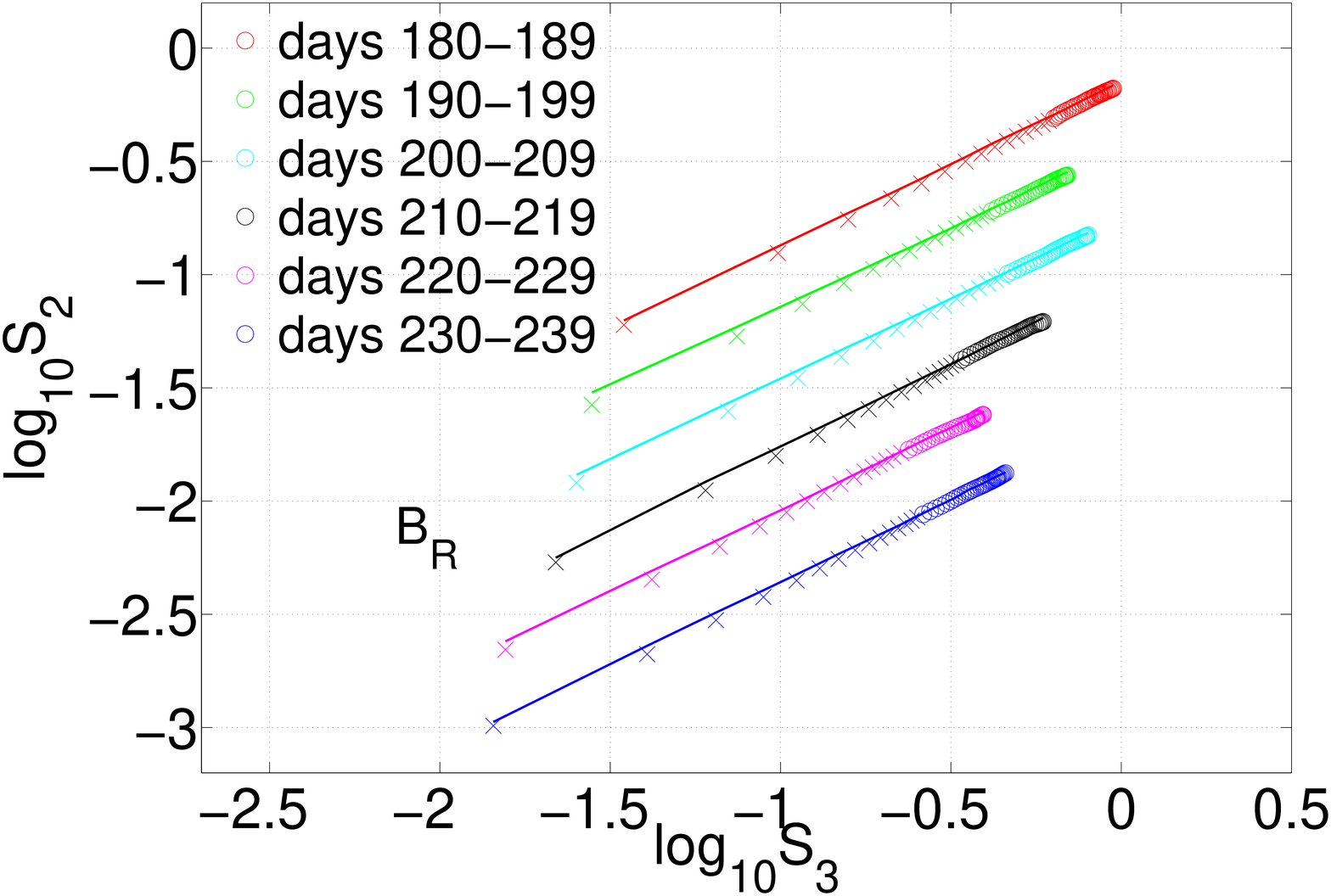}
\plotone{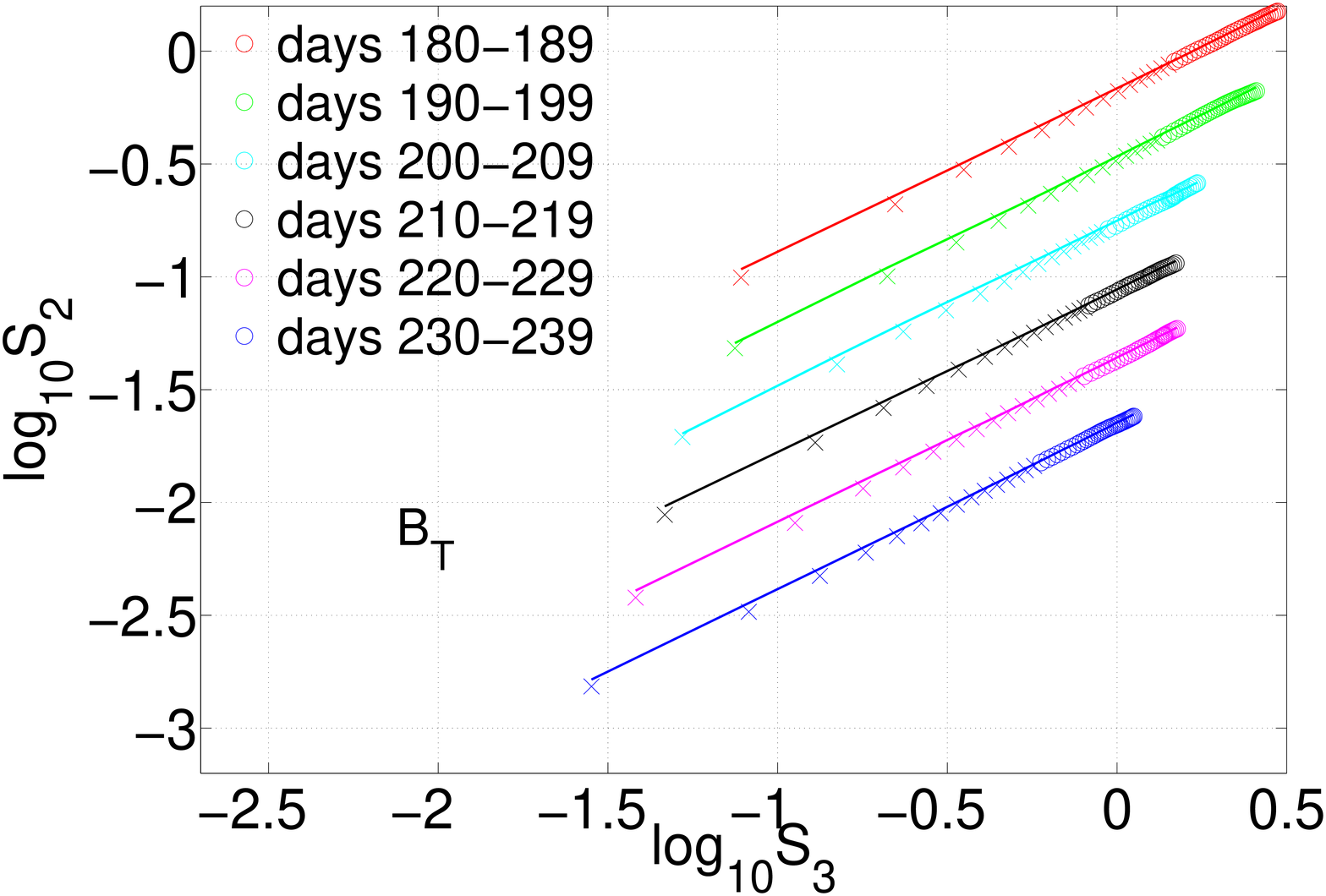}
\plotone{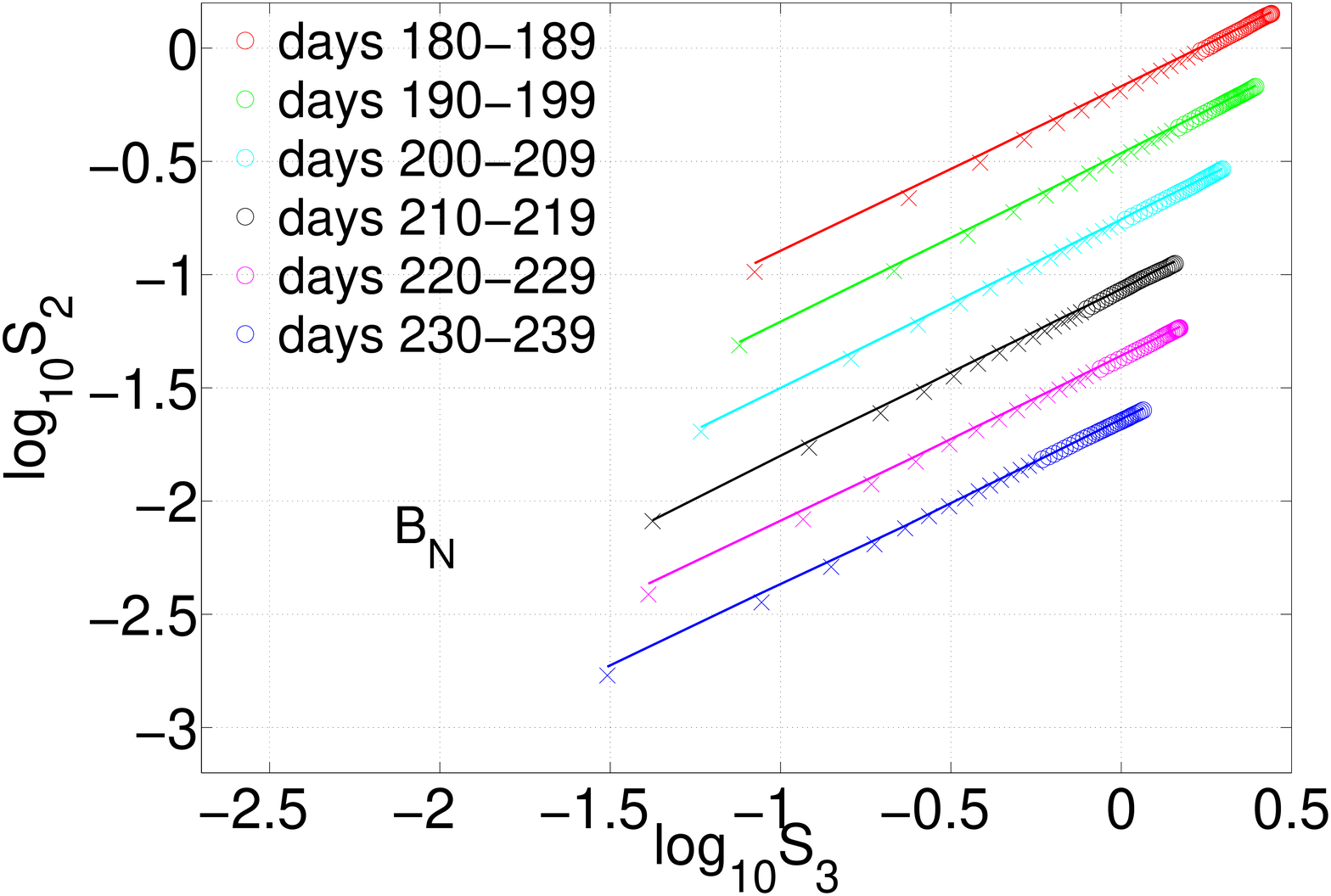}
\caption{Evidence for extended self-similarity (ESS) across the full $\tau$ range. Log-log plots of second order structure function $S_2$ versus third order structure function $S_3$ for all three components of magnetic field fluctuations in the solar wind measured by the ULYSSES spacecraft during contiguous intervals of ten days, which are plotted separately on each panel, from day $180$ to day $239$ of $1995$. The different intervals have been uniformly shifted in the y-direction for clarity. Top left panel: radial field $B_R$. Top right panel: tangential field $B_T$. Bottom panel: normal field $B_N$. Data points in the inertial range are marked by crosses, and in the ``$f^{-1}$'' range by open circles. The straight lines show linear regression fits across the full temporal range from $\tau=2-49$min. These results imply a global average fitting across the different time intervals of $\zeta_R(2)/\zeta_R(3) = 0.749 \pm 0.004$, $\zeta_T(2)/\zeta_T(3) = 0.759 \pm 0.004$ and $\zeta_N(2)/\zeta_N(3) = 0.765 \pm 0.004$.}
\label{Fig.6}
\end{figure}

\begin{figure}
\figurenum{7}
\epsscale{0.8}
\plotone{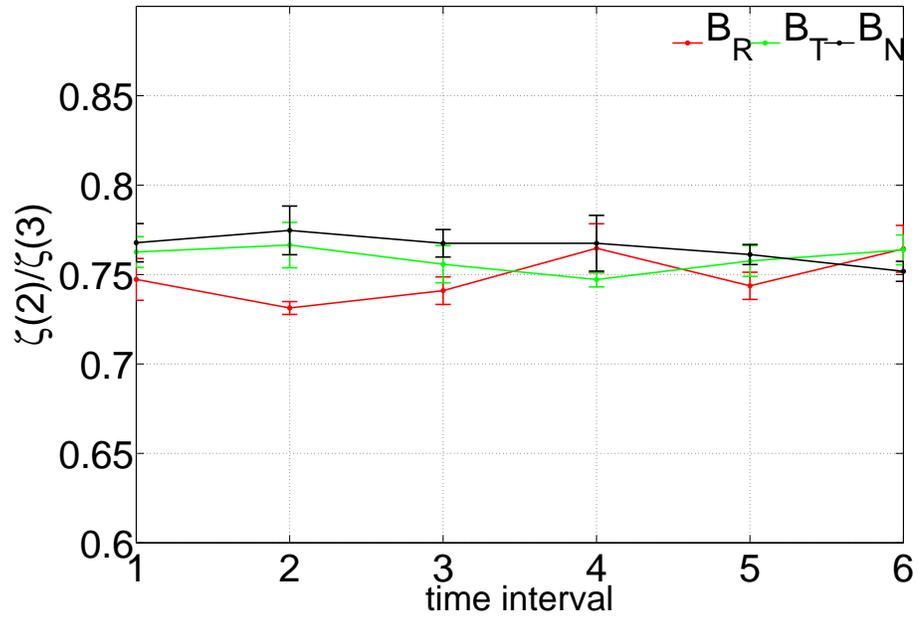}
\caption{Variations of $\zeta(2)/\zeta(3)$ for fits to the full $\tau$ range ($\tau=2-49$min.) for $B_R$, $B_T$ and $B_N$. Here interval one refers to days $180$ to $189$ of $1995$, and interval six to days $230$ to $239$.}
\label{Fig.7}
\end{figure}

\begin{figure} 
\figurenum{8}
\epsscale{0.4}
\plotone{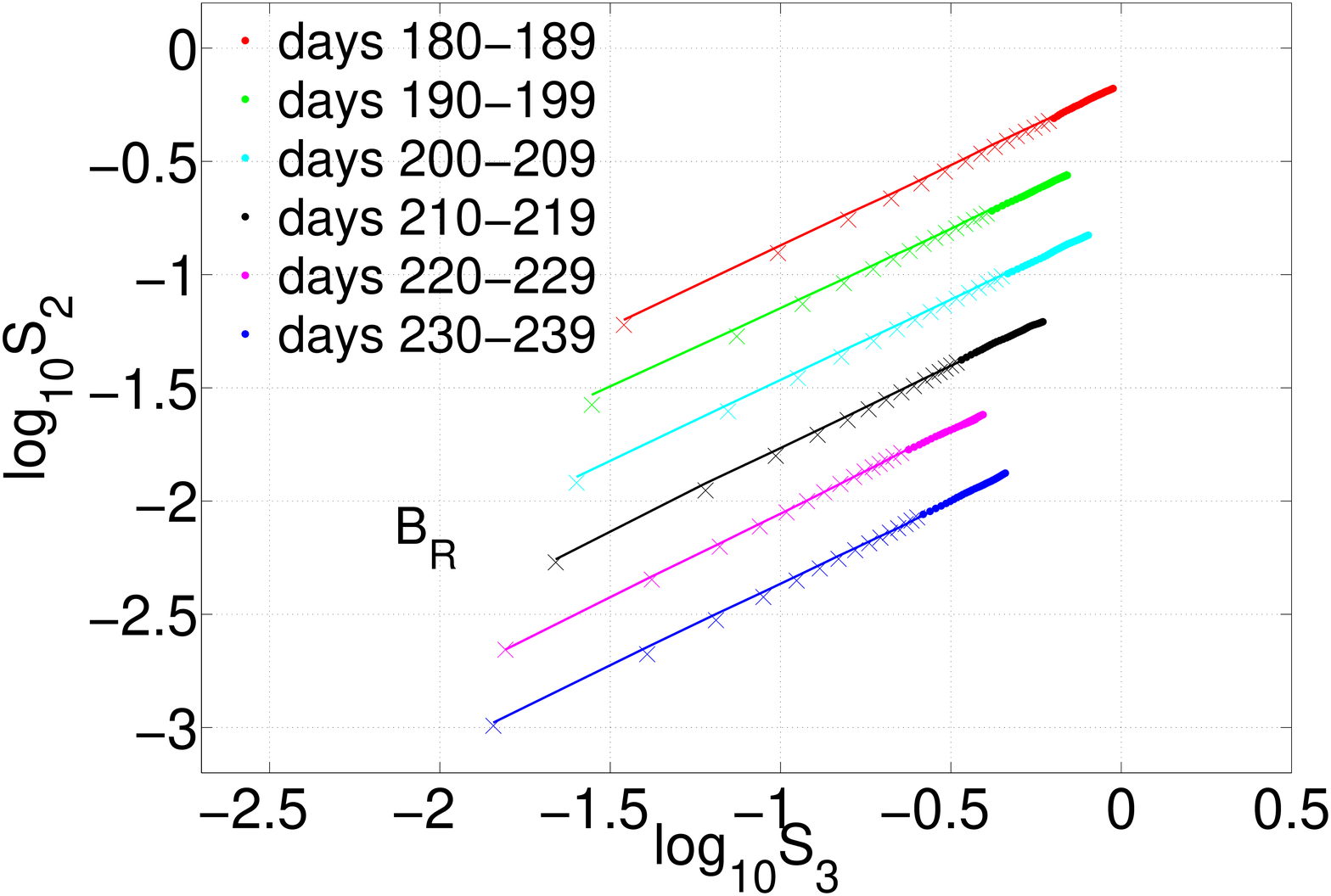}
\plotone{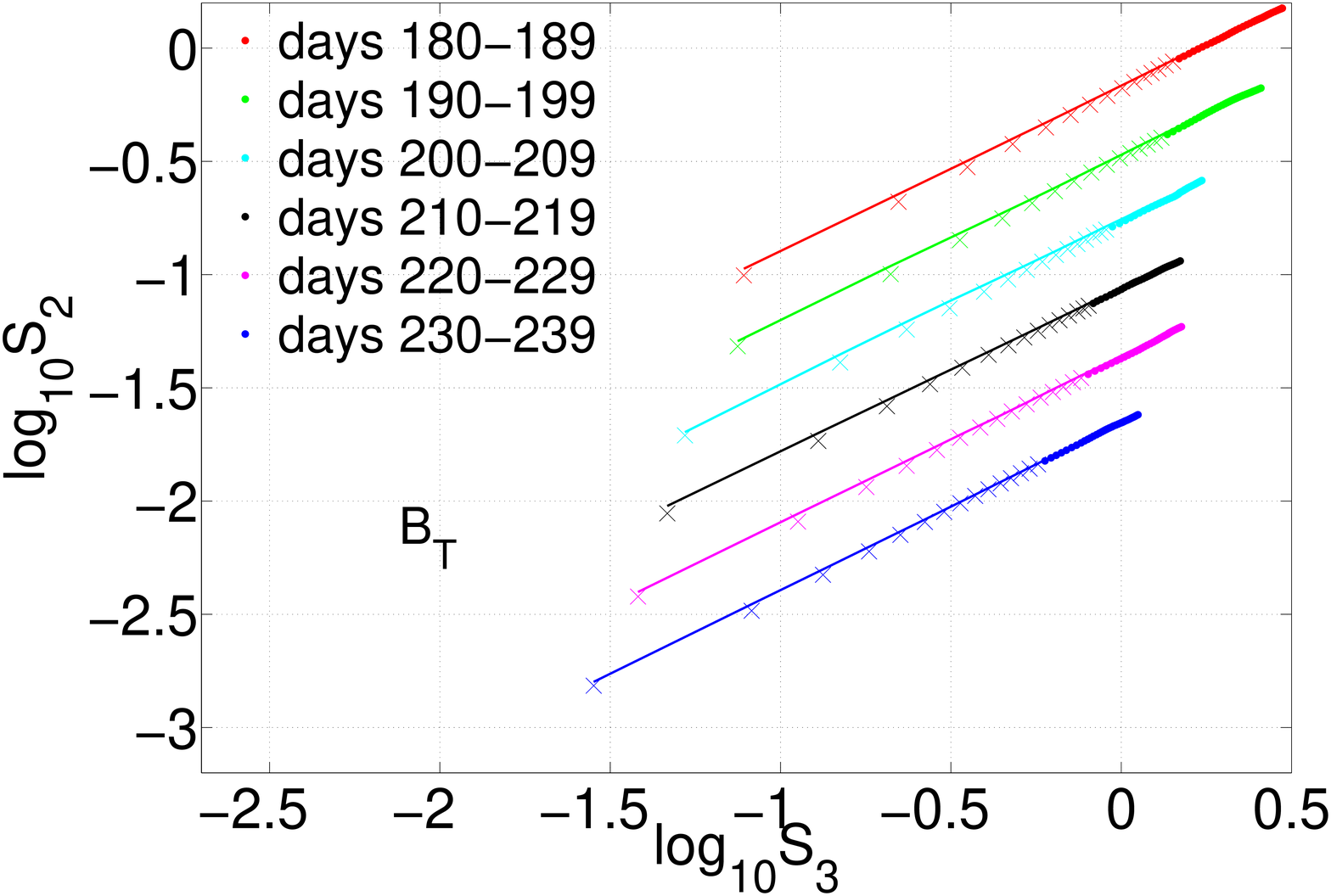}
\plotone{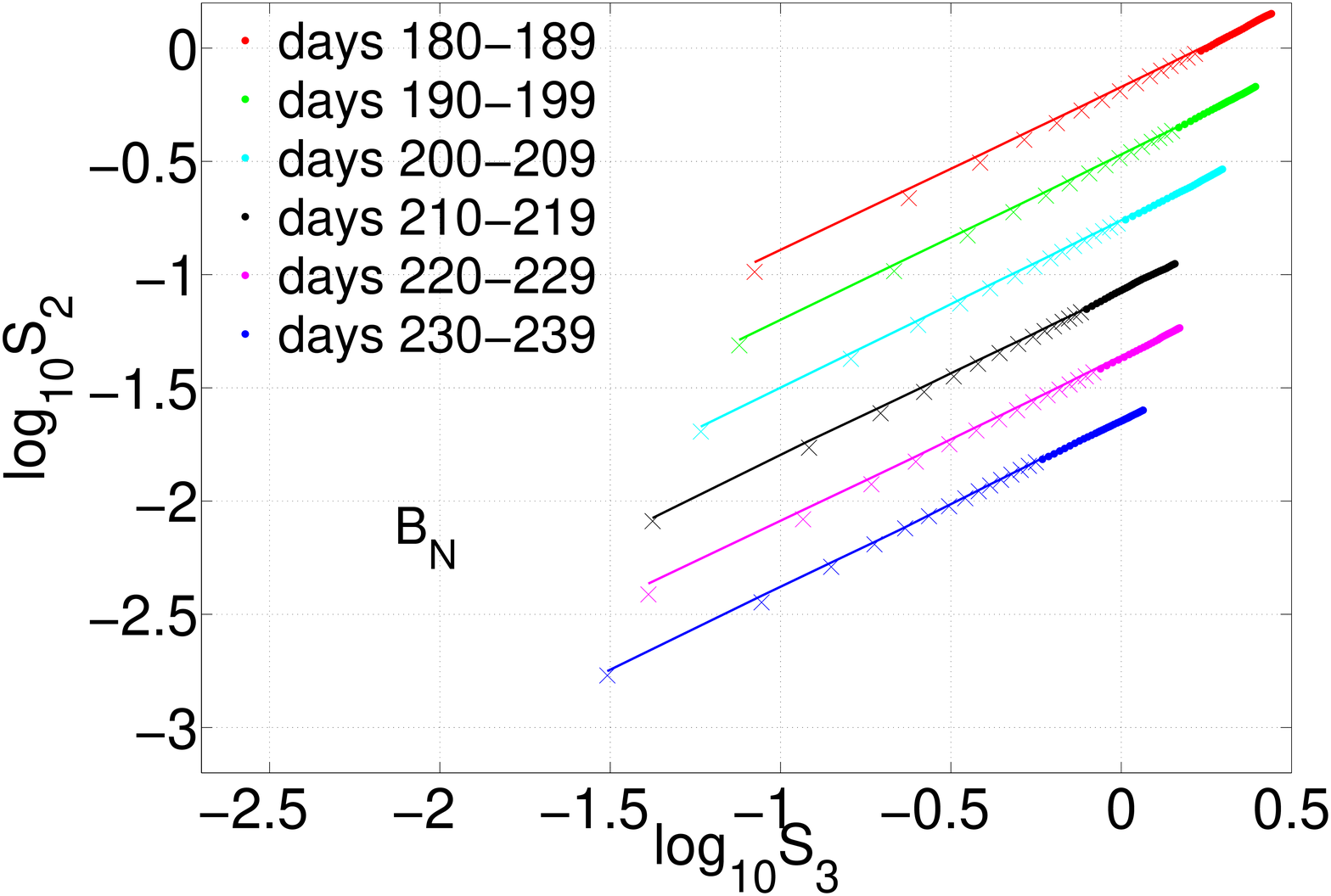}
\caption{Like Figure \ref{Fig.6} but showing fits for the inertial range between $\tau=2-14$ min. These results imply that on average across the different time intervals $\zeta_{S,R}(2)/\zeta_{S,R}(3) = 0.747 \pm 0.008$, $\zeta_{S,T}(2)/\zeta_{S,T}(3) = 0.757 \pm 0.007$ and $\zeta_{S,N}(2)/\zeta_{S,N}(3) = 0.753 \pm 0.005$ in the inertial range.}
\label{Fig.8}
\end{figure}

\begin{figure} 
\figurenum{9}
\epsscale{0.4}
\plotone{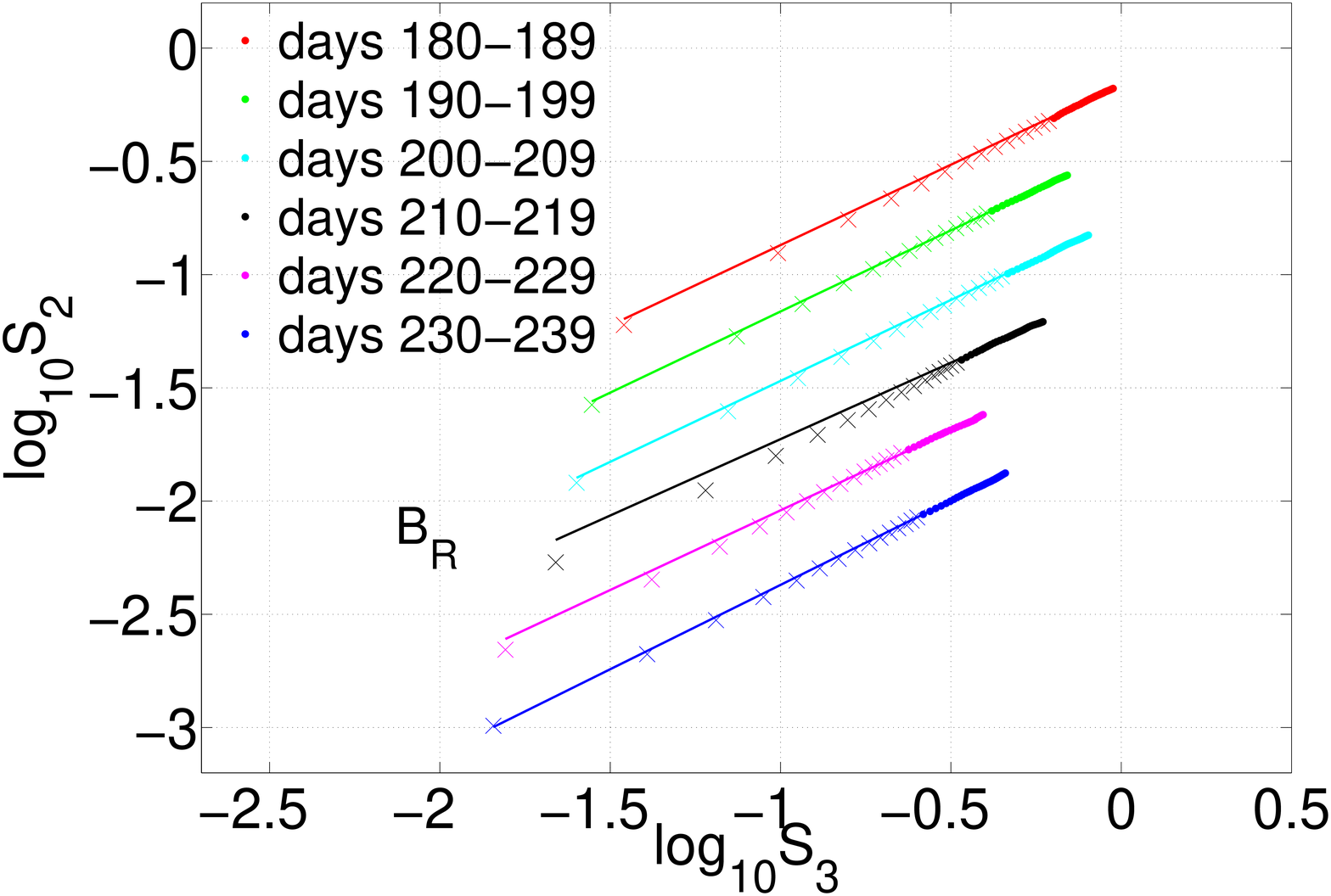}
\plotone{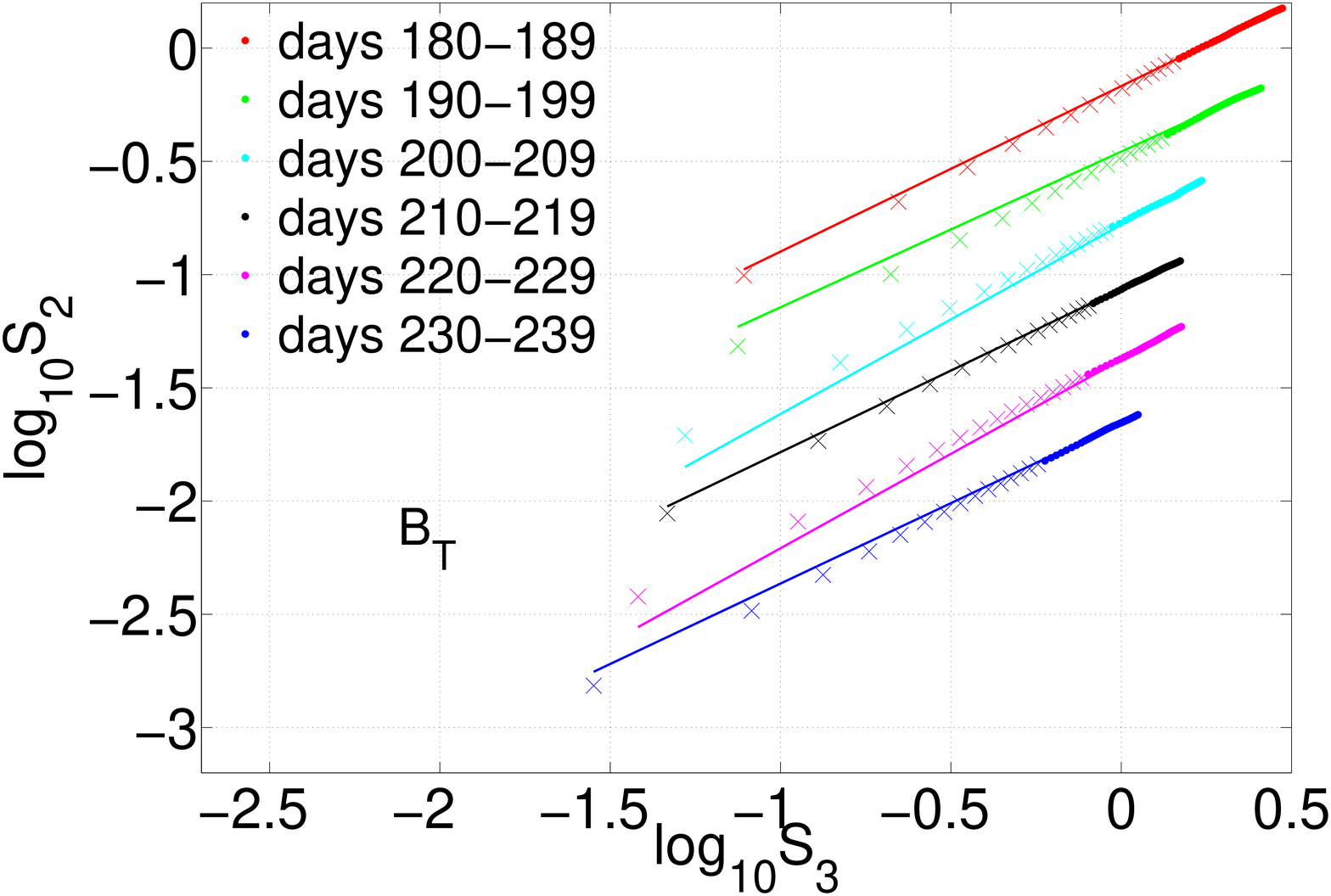}
\plotone{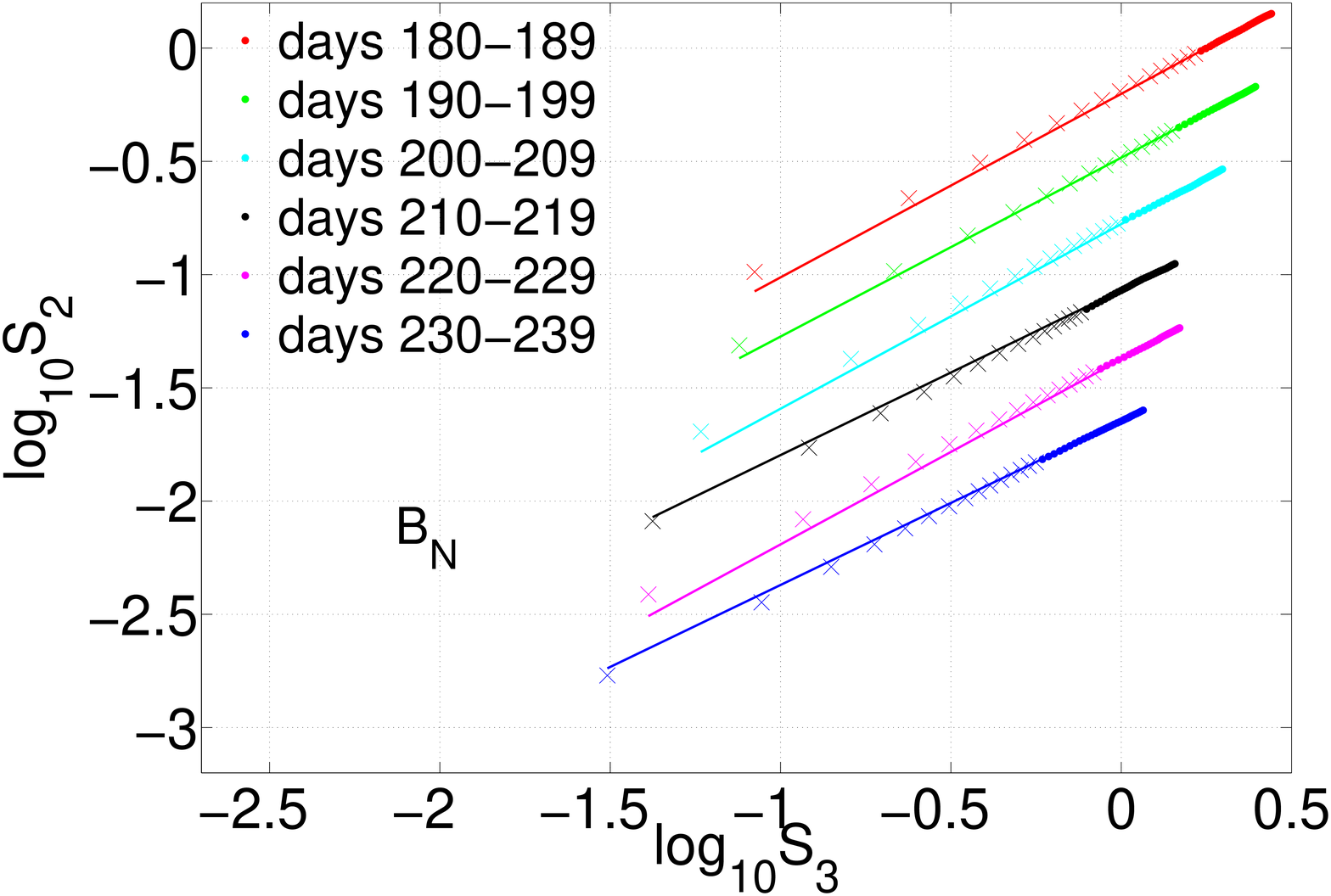}
\caption{Like Figure \ref{Fig.6} but showing fits for the ``$f^{-1}$'' range between $\tau=26-49$ min. These results imply that across the different time intervals $0.675 \pm 0.017 \leq \zeta_{L,R}(2)/\zeta_{L,R}(3)\leq 0.745 \pm 0.017$, $0.687 \pm 0.019 \leq \zeta_{L,T}(2)/\zeta_{L,T}(3) \leq 0.835 \pm 0.034$ and $0.724 \pm 0.004 \leq \zeta_{L,N}(2)/\zeta_{L,N}(3) \leq 0.817 \pm 0.012$ in the ``$f^{-1}$'' range.}
\label{Fig.9}
\end{figure}

\begin{figure}

\figurenum{10}
\epsscale{0.8}
\plotone{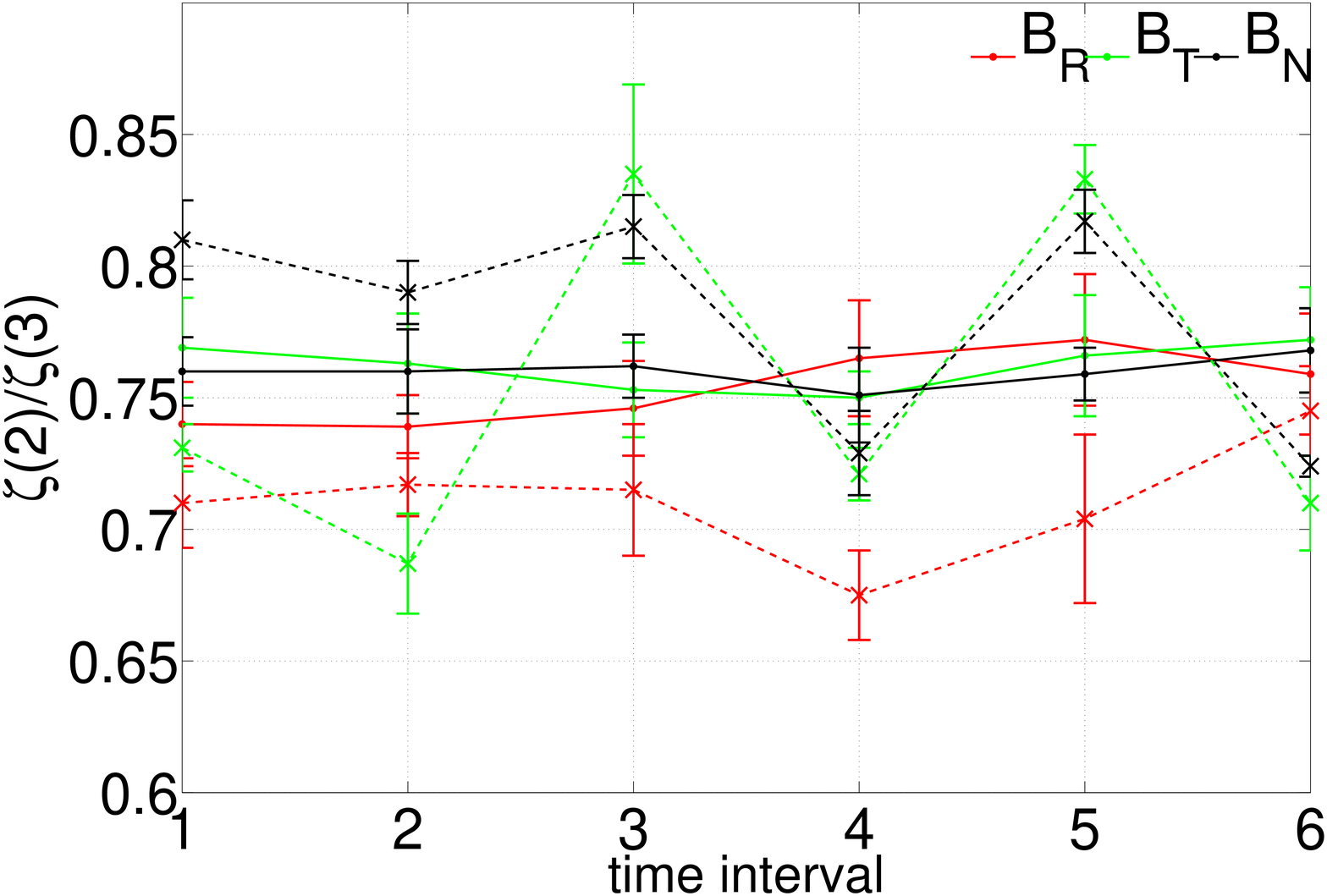}
\caption{Variations of $\zeta(2)/\zeta(3)$ for fits to the inertial range (solid lines, $\tau=2-14$min.) and the ``$f^{-1}$'' range (dashed lines, $\tau=26-49$min.) for $B_R$, $B_T$ and $B_N$. Here interval one refers to days $180$ to $189$ of $1995$, and interval six to days $230$ to $239$.}
\label{Fig.10}
\end{figure}

The difference between the inertial range and ``$f^{-1}$'' fits can be seen in Figure \ref{Fig.10}, where the gradients of the fits are shown. Importantly, Figure \ref{Fig.10} demonstrates that the dominant contribution to the scatter in the corresponding global plot, Figure \ref{Fig.5}, arises from the ``$f^{-1}$'' region at larger $\tau$. This aligns with what was found with the GSF analysis, see Figure \ref{Fig.3}.

\begin{figure}
\figurenum{11}
\epsscale{0.8}
\plotone{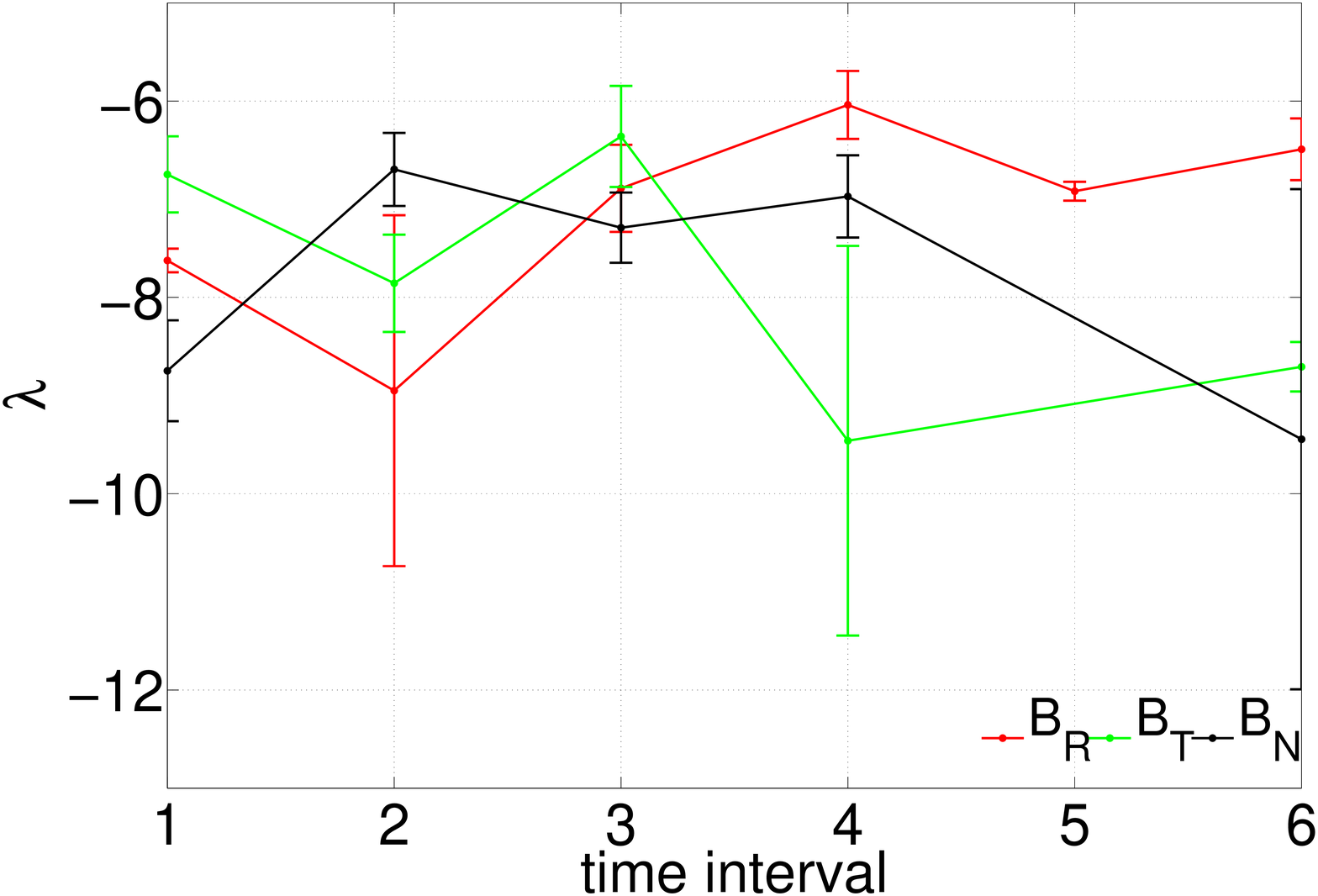}
\caption{Variations of $\lambda$ for $\alpha$ and $\beta$ quadratic fits (see equations (\ref{eqn7}) and (\ref{eqn8})) to the inertial range ($\tau=2-14$min.) for $B_R$, $B_T$ and $B_N$. Here interval one refers to days $180$ to $189$ of $1995$, and interval six to days $230$ to $239$.}
\label{Fig.11}
\end{figure}

\begin{figure} 
\figurenum{12}
\epsscale{0.4}
\plotone{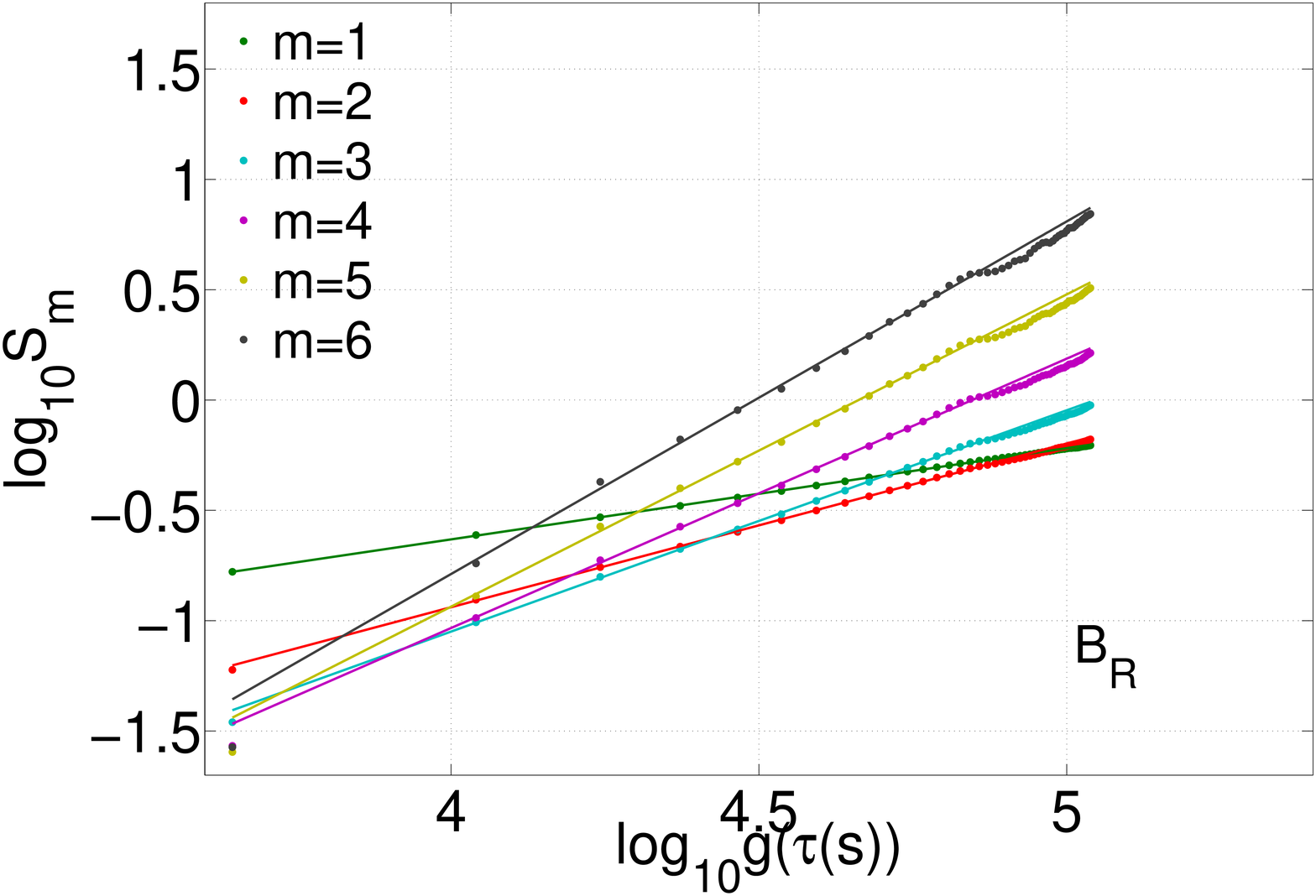}
\plotone{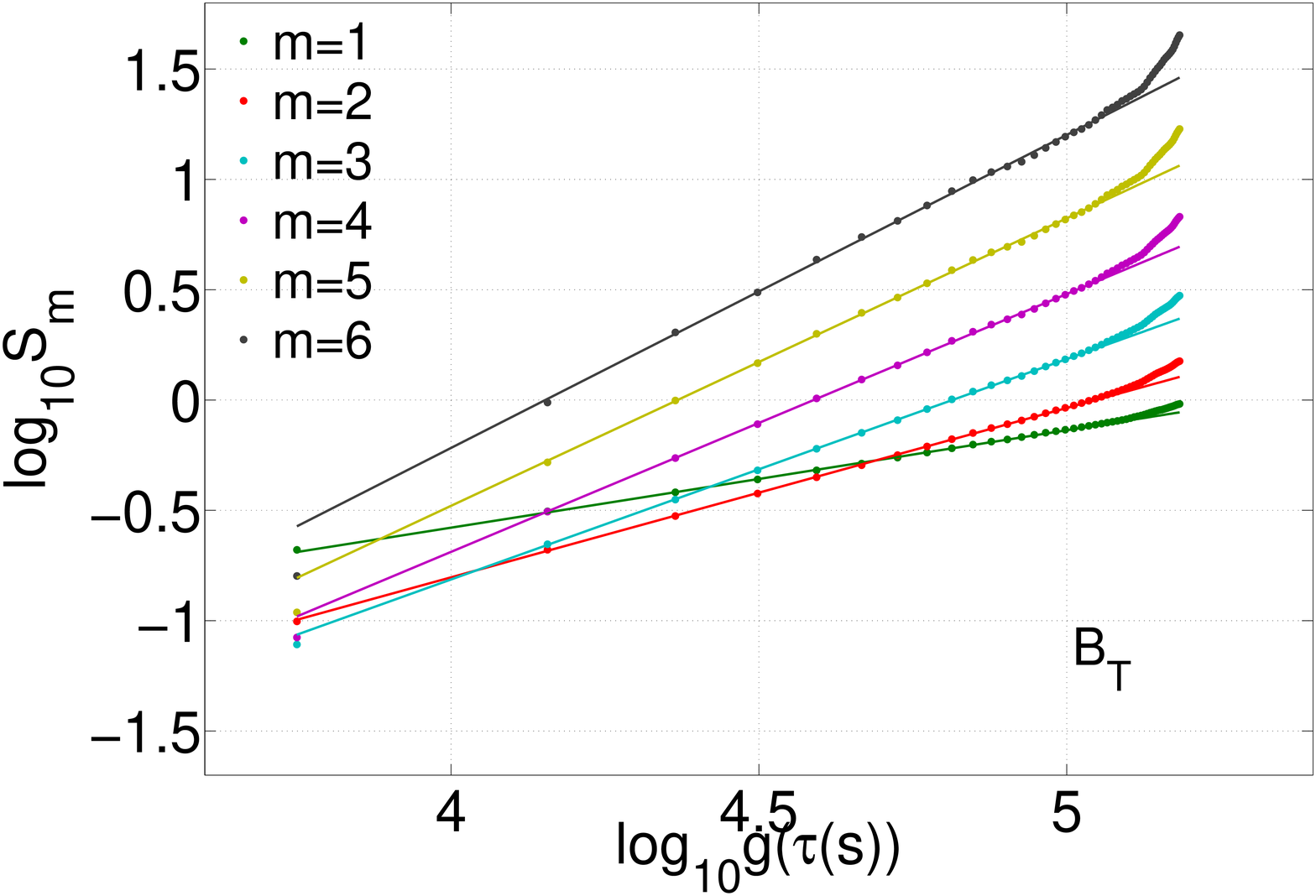}
\plotone{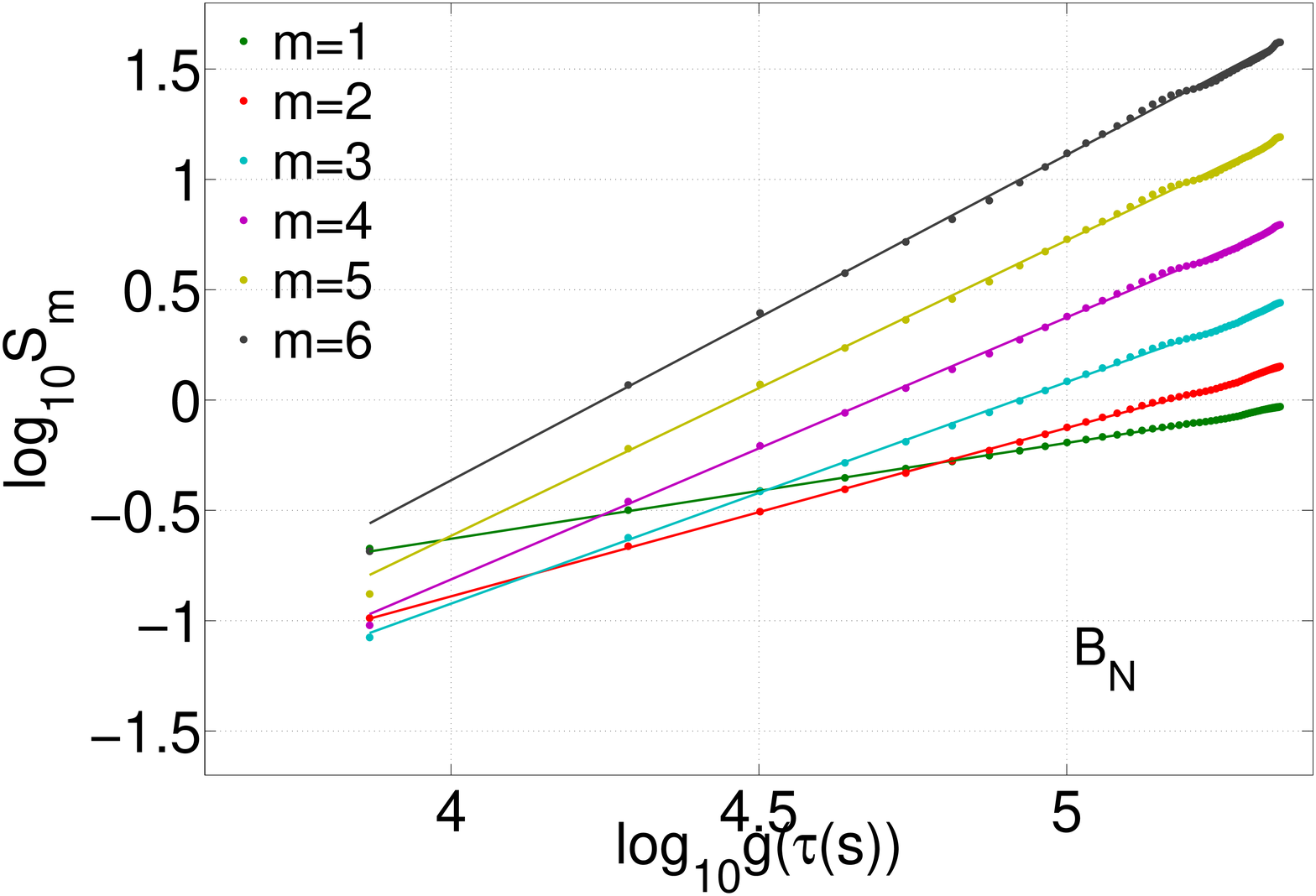}
\caption{Evidence for $g(\tau)$ dependence in the inertial range. Log-log plots of the GSFs for moments $m=0-6$ versus $g(\tau)$ for all three components. A single time interval is shown here, days $180-189$; the same analysis was done for the other time intervals and similar results found. The $g(\tau)$ is normalized for each component such that $\zeta_{S}(3)=1$. This is achieved by obtaining $\zeta_{S}(3)$ from a linear regression fit of the GSF for each component and then incorporating this into the $g(\tau)$ expression such that $g(\tau)=\tau^{-log_{10}(\tilde{\lambda}\tau)\zeta_{i}(3)}$ where $i=R,\;T\;$ or $N$. The break between the inertial range and ``$f^{-1}$'' ranges does not seem to change position from the GSF but is clearer in these plots.}
\label{Fig.12}
\end{figure}

The GSF plots in Figure \ref{Fig.2} can be fitted using a quadratic fit of the form
\begin{equation}
log_{10}S_m(\tau)\sim \alpha(m)(log_{10}\tau)^{2}+\beta(m)(log_{10}\tau)+\gamma \label{eqn7}
\end{equation}
where $\gamma$ contains the secular variation in $S_{3}$ seen in Figure \ref{Fig.5}. Equation (\ref{eqn7}) can be rewritten in the form
\begin{equation}
S_m(\tau)\sim \tau^{-a(m)(log_{10}(\tau)+\lambda)}\sim[\tau^{-log_{10}(\tilde{\lambda}\tau)}]^{a(m)}\label{eqn8}
\end{equation}
where $\tilde{\lambda}=10^{\lambda}=10^{-7.575\pm0.246}$ is found by taking an average over fits for all the different time intervals. The minus sign is necessary if we insist on positive scaling of exponents $a(m)$ as $\alpha$ is negative and $\beta$ is positive. In Figure \ref{Fig.11}, we show $\lambda$ for all components and time intervals. The recovery of a power law behavior in the inertial range can be seen in Figure \ref{Fig.12} by plotting the GSF versus the $g(\tau)$ defined in equation (\ref{eqn8}) rather than $\tau$, where the form of $g(\tau)$ above is normalized such that $\zeta_S(3)=1$.

Finally let us again consider the behavior of the $\zeta_S$ exponents for $m=1-6$ for all components and time intervals. These can be fitted to the multifractal model for fully developed turbulence proposed by \citet{multiPRL87} (p-model). A simple example of this is to consider an eddy decaying into two smaller ones. The parent transfers a fraction $p$ of its energy to one of them and a fraction $(1-p)$ to the other one (by convention $p\geq1/2$). The values of $\zeta(m)$ are given by
\begin{equation}
\zeta(m)=1-log_{2}\left(p^{m/3}+(1-p)^{m/3}\right)\label{eqn9}
\end{equation}
which reduces to Kolmogorov scaling, i.e $\zeta(m)=m/3$, for $p=1/2$. For the data considered here, p-model fits of the $\zeta_S$ exponents for the different components during the first time interval give $p_R=0.79\pm0.01$, $p_T=0.85\pm0.01$ and $p_N=0.84\pm0.01$. These values are consistent with previous work by \citet{MHDcascade97}, who found $p\sim0.8$, and by \citet{magstudyUlysses}.

\section{Conclusions}

In this paper we have analysed ULYSSES quiet fast polar solar wind magnetic field measurements to study the evolving turbulence. We quantify the scaling behavior of both the inertial range (i.e. power spectrum $\sim f^{-5/3}$) and the lower frequency ``$f^{-1}$'' range present in the solar wind. Six contiguous intervals of ten days, over which approximate stationarity can be assumed, were studied using the Taylor hypothesis to relate temporal and spatial scales. The scaling that we establish is ``macroscopic'' in the sense that it is obtained over these time intervals of ten days. One can also consider local (in time) scaling properties \citep{PRLSorriso07}. We use generalized structure functions (GSF) and extended self-similarity (ESS) to quantify statistical scaling, and find that:
\begin{itemize}
\item GSF is sufficient to reveal power law scaling in the low frequency ``$f^{-1}$'' range of the form $\langle\vert y(t+\tau)-y(t)\vert^{m}\rangle\sim\tau^{\zeta(m)}$, but ESS is necessary to reveal scaling in the inertial range. This implies a scaling of the form $\langle\vert y(t+\tau)-y(t)\vert^{m}\rangle\sim [g(\tau)]^{\zeta(m)}$ over the inertial range.
\item The ``$f^{-1}$'' range scaling varies in a non secular way with spacecraft position as found previously \citep{evolvingturbpolar}. This is consistent with a coronal origin for the ``$f^{-1}$'' scaling.
\item In the inertial range, comparisons of the third order structure function $S_{3}$ for the different time intervals show that $g(\tau)$ is independent of spacecraft position, although an ordered trend in the absolute value of $S_{3}$ with increasing radial distance is observed.
\item A good fit to the inertial range is $g(\tau)\sim\tau^{-log_{10}(\tilde{\lambda}\tau)}$, where $\tilde{\lambda}=10^{-7.575\pm0.246}$.
\item The exponents found for the inertial range, normalized such that $\zeta_S(3)=1$, are fitted by a p-model with $p_R=0.79\pm0.01$, $p_T=0.85\pm0.01$ and $p_N=0.84\pm0.01$. This implies a higher degree of intermittency in the normal components of the magnetic field than in the radial component.
\end{itemize}
Our results clearly differentiate between the dynamics of the fluctuations seen in the ``$f^{-1}$'' and in the inertial range. Intriguingly, the inertial range signature is not simply that expected from homogeneous turbulence, in that there is a robust ``envelope'' function $g(\tau)$ for the scaling. This is highly suggestive of turbulence in a confined or space-varying medium, see for example the work of \citet{biskampsimul}. Our function $g(\tau)$ may therefore capture the evolution of the turbulence observed at ULYSSES, reflecting both the heating of the fast solar wind at the corona and the subsequent expansion in the presence of the large scale solar magnetic field.
\acknowledgments
We thank the National Space Science Data Center and the Principal Investigator, A. Balogh, for this use of the ULYSSES magnetometer data. R. M. N. acknowledges the support of the STFC and UKAEA.

\clearpage


\begin{thebibliography}{}
\bibitem[Bak et al.(1987)]{1overfnoise} Bak, P., Tang, C., \& Wiesenfeld, K.\ 1987, \prl, 59, 381 
\bibitem[Balogh et al.(1992)]{instrumentULYSSES} Balogh, A., Beek, T.~J., Forsyth, R.~J., Hedgecock, P.~C., Marquedant, R.~J., Smith, E.~J., Southwood, D.~J., \& Tsurutani, B.~T.\ 1992, \aaps, 92, 221 
\bibitem[Bame et al.(1992)]{SWOOPS} Bame, S.~J., McComas, D.~J., Barraclough, B.~L., Phillips, J.~L., Sofaly, K.~J., Chavez, J.~C., Goldstein, B.~E., \& Sakurai, R.~K.\ 1992, \aaps, 92, 237 
\bibitem[Bavassano et al.(1982)]{Bav1982} Bavassano, B., Dobrowolny, M., Mariani, F., \& Ness, N.~F.\ 1982, \jgr, 87, 3617 
\bibitem[Benzi et al.(1993)]{ESS93} Benzi, R., Ciliberto, S., Tripiccione, R., Baudet, C., Massaioli, F., \& Succi, S.\ 1993, \pre, 48, 29 
\bibitem[Biskamp \& M{\H u}ller(2000)]{biskampsimul} Biskamp, D. M{\H u}ller, W.-C.\ 2000, Physics of Plasmas, 7, 4889 
\bibitem[Bruno et al.(2003)]{radialintermittency} Bruno, R., Carbone, V., Sorriso-Valvo, L., \& Bavassano, B.\ 2003, Journal of Geophysical Research (Space Physics), 108, 1130 
\bibitem[Bruno \& Carbone(2005)]{livingreview} Bruno, R., \& Carbone, V.\ 2005, Living Reviews in Solar Physics, 2, 4 
\bibitem[Bruno et al.(2007)]{interimf} Bruno, R., Carbone, V., Chapman, S., Hnat, B., Noullez, A., \& Sorriso-Valvo, L.\ 2007, Physics of Plasmas, 14, 2901
\bibitem[Burlaga \& Klein(1986)]{fractalBurlaga} Burlaga, L.~F., \& Klein, L.~W.\ 1986, \jgr, 91, 347 
\bibitem[Carbone et al.(1996)]{ESSCarbone} Carbone, V., Bruno, R., \& Veltri, P.\ 1996, \grl, 23, 121 
\bibitem[Chapman et al.(2005)]{scaling05} Chapman, S.~C., Hnat, B., Rowlands, G., \& Watkins, N.~W.\ 2005, Nonlinear Processes in Geophysics, 12, 767 
\bibitem[Coleman(1968)]{kolmogorov68} Coleman, P.~J., Jr.\ 1968, \apj, 153, 371  
\bibitem[Crosby et al.(1993)]{solarflare} Crosby, N., Aschwanden, M., \& Dennis, B.\ 1993, Advances in Space Research, 13, 179 
\bibitem[Dendy \& Chapman(2006)]{characternonlinear} Dendy, R.~O., \& Chapman, S.~C.\ 2006, Plasma Physics and Controlled Fusion, 48, B313 
\bibitem[Dudson et al.(2005)]{L_H_mast} Dudson, B.~D., Dendy, R.~O., Kirk, A., Meyer, H., \& Counsell, G.~F.\ 2005, Plasma Physics and Controlled Fusion, 47, 885 
\bibitem[Erd{\H o}s \& Balogh(2005)]{insitub} Erd{\H o}s, G., \& Balogh, A.\ 2005, Advances in Space Research, 35, 625  
\bibitem[Feynman et al.(1995)]{breakpoint} Feynman, J., Ruzmaikin, A., \& Smith, E.~J.\ 1995, Solar Wind Conference, 80 
\bibitem[Forsyth et al.(1996)]{solarmin} Forsyth, R.~J., Balogh, A., Horbury, T.~S., Erd{\H o}s, G., Smith, E.~J., \& Burton, M.~E.\ 1996, \aap, 316, 287 
\bibitem[Frisch(1995)]{Frisch} Frisch, U.\ 1995, Turbulence : the legacy of A. N. Kolmogorov, 2nd ed.~by Uriel Frisch.~ Cambridge University Press.~Cambridge, 1995, 
\bibitem[Goldstein et al.(1995a)]{reviewMHDinSW} Goldstein, M.~L., Roberts, D.~A., \& Matthaeus, W.~H.\ 1995, \araa, 33, 283 
\bibitem[Goldstein et al.(1995b)]{mhd_high_lat} Goldstein, B.~E., Smith, E.~J., Balogh, A., Horbury, T.~S., Goldstein, M.~L., \& Roberts, D.~A.\ 1995, \grl, 22, 3393 
\bibitem[Goldstein(2001)]{majprobspp} Goldstein, M.~L.\ 2001, \apss, 277, 349 
\bibitem[Grossmann et al.(1997)]{ESSGross} Grossmann, S., Lohse, D., \& Reeh, A.\ 1997, \pre, 56, 5473 
\bibitem[Horbury et al.(1995a)]{secminKol} Horbury, T., Balogh, A., Forsyth, R.~J., \& Smith, E.~J.\ 1995a, Annales Geophysicae, 13, 105 
\bibitem[Horbury et al.(1995b)]{evolvingturbpolar} Horbury, T.~S., Balogh, A., Forsyth, R.~J., \& Smith, E.~J.\ 1995b, \grl, 22, 3401 
\bibitem[Horbury et al.(1996a)]{turbulentevolution} Horbury, T.~S., Balogh, A., Forsyth, R.~J., \& Smith, E.~J.\ 1996a, \jgr, 101, 405 
\bibitem[Horbury et al.(1996b)]{magsignature} Horbury, T.~S., Balogh, A., Forsyth, R.~J., \& Smith, E.~J.\ 1996b, \aap, 316, 333 
\bibitem[Horbury \& Balogh(1997)]{MHDcascade97} Horbury, T.~S., \& Balogh, A.\ 1997, Nonlinear Processes in Geophysics, 4, 185
\bibitem[Horbury \& Balogh(2001)]{Evolutionmag} Horbury, T.~S., \& Balogh, A.\ 2001, \jgr, 106, 15929 
\bibitem[Kiyani et al.(2006)]{scalingmethod06} Kiyani, K., Chapman, S.~C., \& Hnat, B.\ 2006, \pre, 74, 051122 
\bibitem[Kiyani et al.(2007)]{selfsimilarwind} Kiyani, K., Chapman, S.~C., Hnat, B., \& Nicol, R.~M.\ 2007, \prl, 98, 211101 
\bibitem[Kolmogorov(1941)]{K41} Kolmogorov, A.\ 1941, Akademiia Nauk SSSR Doklady, 30, 301, This translation by Levin, V.
\bibitem[Marsch \& Tu(1996)]{spatialspectralexponent} Marsch, E., \& Tu, C.-Y.\ 1996, \jgr, 101, 11149 
\bibitem[Matthaeus \& Goldstein(1986)]{PRL1/f86} Matthaeus, W.~H., \& Goldstein, M.~L.\ 1986, \prl, 57, 495 
\bibitem[Matthaeus et al.(2005)]{Matt2point} Matthaeus, W.~H., Dasso, S., Weygand, J.~M., Milano, L.~J., Smith, C.~W., \& Kivelson, M.~G.\ 2005, \prl, 95, 231101 
\bibitem[Matthaeus et al.(2007)]{densityandB1overf} Matthaeus, W.~H., Breech, B., Dmitruk, P., Bemporad, A., Poletto, G., Velli, M., \& Romoli, M.\ 2007, \apjl, 657, L121 
\bibitem[Meneveau \& Sreenivasan(1987)]{multiPRL87} Meneveau, C., \& Sreenivasan, K.~R.\ 1987, \prl, 59, 1424 
\bibitem[Merrifield et al.(2007)]{simulmerrifield} Merrifield, J.~A., Chapman, S.~C., \& Dendy, R.~O.\ 2007, Phys. Plasmas, 14, 012301 
\bibitem[Ofman(2005)]{coronalheating} Ofman, L.\ 2005, Space Science Reviews, 120, 67 
\bibitem[Pagel \& Balogh(2001)]{magstudyUlysses} Pagel, C., \& Balogh, A.\ 2001, Nonlinear Processes in Geophysics, 8, 313 
\bibitem[Percival \& Walden(1993)]{pmtm} Percival, D.B., \& A.T. Walden, Spectral Analysis for Physical Applications: Multitaper and Conventional Univariate Techniques, Cambridge University Press, 1993.
\bibitem[Phillips et al.(1995)]{plasmaobservations} Phillips, J.~L., et al.\ 1995, Science, 268, 1030 
\bibitem[Ruzmaikin et al.(1993)]{fractalRuz} Ruzmaikin, A., Lyannaya, I.~P., Styashkin, V.~A., \& Eroshenko, E.\ 1993, \jgr, 98, 13303 
\bibitem[Ruzmaikin et al.(1995a)]{intermittentturbsouth} Ruzmaikin, A.~A., Feynman, J., Goldstein, B.~E., Smith, E.~J., \& Balogh, A.\ 1995a, \jgr, 100, 3395 
\bibitem[Ruzmaikin et al.(1995b)]{confAIP} Ruzmaikin, A., Goldstein, B.~E., Smith, E.~J., \& Balogh, A.\ 1995b, Solar Wind Conference, 38 
\bibitem[Schrijver et al.(1992)]{mag_carpet_2} Schrijver, C.~J., Zwaan, C., Balke, A.~C., Tarbell, T.~D., \& Lawrence, J.~K.\ 1992, \aap, 253, L1 
\bibitem[Smith et al.(1995)]{ulysseslatgradient} Smith, E.~J., Balogh, A., Lepping, R.~P., Neugebauer, M., Phillips, J., \& Tsurutani, B.~T.\ 1995, Advances in Space Research, 16, 165 
\bibitem[Smith \& Balogh(1995)]{invariantBr} Smith, E.~J., \& Balogh, A.\ 1995, \grl, 22, 3317 
\bibitem[Smith et al.(1995)]{polaralfven} Smith, E.~J., Balogh, A., Neugebauer, M., \& McComas, D.\ 1995, \grl, 22, 3381 
\bibitem[Sorriso-Valvo et al.(2007)]{PRLSorriso07} Sorriso-Valvo, L., et al.\ 2007, \prl, 99, 115001
\bibitem[Sornette(2004)]{Sornette} Sornette, D.\ 2004, Critical phenomena in natural sciences : chaos, fractals, selforganization and disorder : concepts and tools, 2nd ed.~by Didier Sornette.~ Springer series in synergetics.~Heidelberg: Springer, 2004, 
\bibitem[Taylor(1938)]{Taylor38} Taylor, G.~I.\ 1938, Royal Society of London Proceedings Series A, 164, 476
\end{thebibliography}
\end{document}